\documentclass[journal]{IEEEtran}

\ifCLASSINFOpdf
\else
\fi
\usepackage{amsmath}
\makeatletter
\renewcommand{\maketag@@@}[1]{\hbox{\m@th\normalsize\normalfont#1}}%
\makeatother
\usepackage{multirow}
\usepackage[pagebackref,pagebackref,colorlinks,linkcolor=red]{hyperref}
\usepackage{markdown}

\usepackage{amsmath}
\usepackage{amsfonts}
\usepackage{array}
\usepackage{multirow}
\usepackage{algorithmic}
\usepackage{array}

\usepackage{graphicx}
\begin{document}

\title{4D~LUT: Learnable Context-Aware 4D Lookup Table for Image Enhancement}

\author{Chengxu~Liu,
        Huan~Yang,
        Jianlong~Fu, 
        Xueming~Qian
    \thanks{This work was done while Chengxu Liu was a research intern at Microsoft Research Asia.}
	\thanks{Chengxu Liu is with the School of Information and Communication Engineering, Xi’an Jiaotong University, Xi’an 710049, China (e-mail: liuchx97@gmail.com).}
	\thanks{Huan Yang and Jianlong Fu are with  Microsoft Research (e-mail: huayan@microsoft.com; jianf@microsoft.com).}
	\thanks{Xueming Qian is with the Ministry of Education Key Laboratory for Intelligent Networks and Network Security, School of Information and Communication Engineering, and SMILES LAB, Xi’an Jiaotong University, Xi’an 710049, China. (*Corresponding author, qianxm@mail.xjtu.edu.cn).}

}

\maketitle

\begin{abstract}
Image enhancement aims at improving the aesthetic visual quality of photos by retouching the color and tone, and is an essential technology for professional digital photography.
Recent years deep learning-based image enhancement algorithms have achieved promising performance and attracted increasing popularity. 
However, typical efforts attempt to construct a uniform enhancer for all pixels' color transformation.
It ignores the pixel differences between different content~(\emph{e.g.,}~sky, ocean, etc.) that are significant for photographs, causing unsatisfactory results.
In this paper, we propose a novel learnable context-aware 4-dimensional lookup table (4D~LUT), which achieves content-dependent enhancement of different contents in each image via adaptively learning of photo context. 
In particular, we first introduce a lightweight context encoder and a parameter encoder to learn a context map for the pixel-level category and a group of image-adaptive coefficients, respectively. Then, the context-aware 4D~LUT is generated by integrating multiple basis 4D~LUTs via the coefficients. Finally, the enhanced image can be obtained by feeding the source image and context map into fused context-aware 4D~LUT via quadrilinear interpolation. 
Compared with traditional 3D~LUT, \emph{i.e.,}~RGB mapping to RGB, which is usually used in camera imaging pipeline systems or tools, 4D LUT, \emph{i.e.,}~RGBC(RGB+Context) mapping to RGB, enables finer control of color transformations for pixels with different content in each image, even though they have the same RGB values.
Experimental results demonstrate that our method outperforms other state-of-the-art methods in widely-used benchmarks.

\end{abstract}


\IEEEpeerreviewmaketitle

\section{Introduction}

\IEEEPARstart{R}{ecent} developments of high precision sensor equipment witness a fast evolution in low-level computer vision fields and digital photography. However, the captured digital photographs still suffer from low quality due to the effects of illumination, weather, camera sensor, photographer skill, and other factors. Image enhancement as an image processing technique to improve the color, contrast, saturation, brightness, and dynamic range can significantly improve the aesthetic visual quality of photos. Compared with manual photo retouching without professional skills and experience, image enhancement algorithms can automatically produce visual-pleasing photos that satisfy visual aesthetics. It can be equipped in smartphones, digital single lens reflex~(DSLR) cameras, and professional-grade software~(\emph{e.g.,}~Photoshop, Lightroom) to provide expert retouching results and has widely promising applications~\cite{liang2021cameranet,qi2021comprehensive,schwartz2018deepisp}.

\begin{figure}
  \centering
  \includegraphics[width = 0.49\textwidth,page=1]{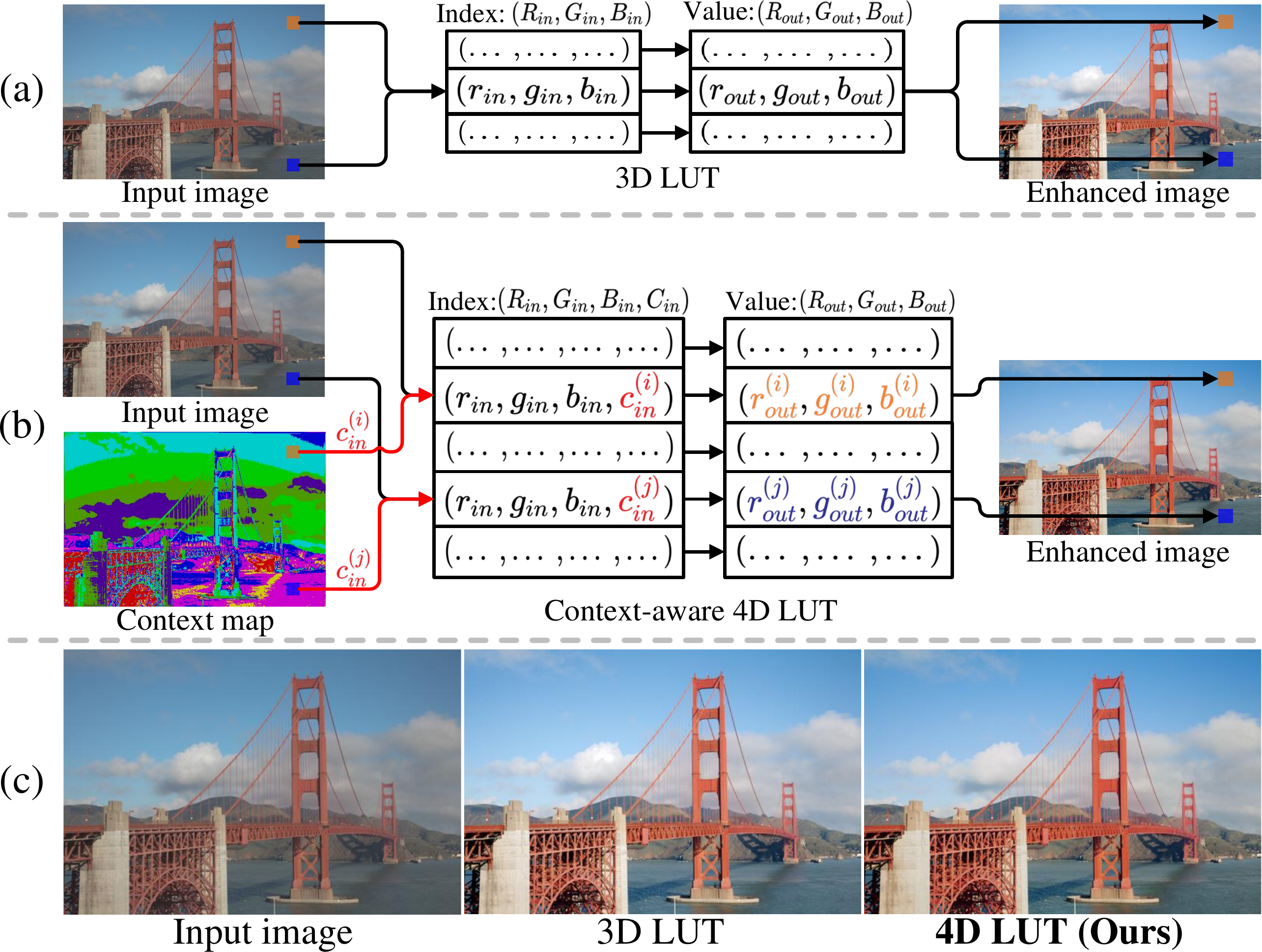}
  \caption{A comparison between 4D~LUT and 3D~LUT~\cite{zeng2020learning}. (a) and (b) indicate the illustration of the 3D~LUT and 4D~LUT, respectively. The enhanced pixels (indicated by \textcolor[RGB]{234,112,14}{orange} and \textcolor{blue}{blue}) are retrieved from the index-value correspondence of the defined LUT according to the original input pixels and context (indicated by \textcolor{red}{red}). (c) shows a result comparison.  They demonstrate that 4D~LUT achieves better results by introducing an additional dimension for content-dependent image enhancement.}
  \label{teaser}
\end{figure}

Traditional image enhancement methods adopt hand-crafted descriptors or filters to adjust the visual quality of an image by feeding a low-quality input image. The hand-crafted global descriptors (\emph{e.g.,}~histogram equalization~\cite{xu2013generalized}, color correction~\cite{lu2015spatio}, etc.) can only be used to roughly change the tone of the whole image by establishing color mapping relationships, while the selectable local filters (\emph{e.g.,}~Laplacian filter~\cite{aubry2014fast}, Guided filter~\cite{he2012guided}, etc.) can finely adjust the visual quality of the image according to the content differences. For example, pixels belonging to natural landscapes, portraits, ancient architecture, etc. should adopt different local filters according to the differences in their content. However, this traditional manual adjustment depends on professional image retouching skills, and retouching the image pixel-by-pixel is time-consuming and not practical enough.

Benefiting from advances in deep learning, learning-based image enhancement algorithms are gaining rapid development~\cite{gharbi2017deep,yan2016automatic}. MIT-Adobe FiveK dataset~\cite{bychkovsky2011learning} proposes an experts-retouched dataset containing 5,000 image pairs of natural landscapes, which is the first to establish a benchmark for the entire field. Further, to facilitate this important and high-visibility task, PPR10K~\cite{liang2021ppr10k} proposes a larger-scale portrait photo retouching dataset, where each portrait has been retouched by three experts with professional experience, separately.

Typical learning-based algorithms can be categorized into three main paradigms. Reinforcement learning-based methods~\cite{hu2018exposure,park2018distort}, image-to-image translation methods~\cite{chen2018deep,he2020conditional,liu2022very}, and physical modeling-based methods~\cite{gharbi2017deep,wang2019underexposed,moran2020deeplpf,moran2021curl,zeng2020learning}. 
Among them, the reinforcement learning-based methods~\cite{hu2018exposure,park2018distort} attempt to decouple the whole process and enhance the image through a step-wise retouching process. 
The image-to-image translation methods~\cite{chen2018deep,he2020conditional,liu2022very} try to establish a mapping from the input to the enhanced output directly through a neural network. That method can yield globally optimized results with an end-to-end training manner. 
However, those algorithms lack interpretability and reliability, like a ``black box", and they also consume lots of computational costs, sacrificing their effectiveness in practical applications. 
To address the limitations of those existing methods, another category of physical modeling-based methods~\cite{gharbi2017deep,wang2019underexposed,moran2020deeplpf,moran2021curl,zeng2020learning} attempt to enhance images using human-interpretable physical models (\emph{e.g.,}~Retinex theory~\cite{land1977retinex}, bilateral filtering~\cite{tomasi1998bilateral}, etc). These methods usually adopt a two-step solution, which includes 1) predicting the relevant physics coefficients based on the proposed physical model and assumptions, and 2) adjusting the original pixels to form an enhanced image through physical theory. 
These efforts not only fail to distinguish the physics coefficients of different content transformations, but also make it difficult to learn in an end-to-end training manner.
Therefore, the physical model-based methods do not provide sufficient enhancement capability.

Recently, some outstanding works~\cite{liang2021ppr10k,zeng2020learning,wang2021real} propose to enhance images by improving the fundamental physical model 3D lookup tables in digital image processing. These methods try to spend fewer runtimes on the enhancement process and focus on learning a uniform enhancer to achieve a globally overall average over all regions of the enhanced result. 
Although the enhanced results can obtain in real-time, they lack the ability to finely control the color transformation of pixels with different content in each image, and can only obtain globally sub-optimal results. 
This issue significantly limits the color richness of enhanced images and cannot be handled by 3D lookup tables.
For example, as shown in Fig.~\ref{teaser}(a), the sky (indicated by orange) and the sea (indicated by blue) in the image (even with the same RGB values) should be adjusted with different transformations to become bluer and greener, respectively, instead of being treated equally.
Intuitively, the content of natural landscapes and portraits should have high color contrast and luminance, respectively, while ancient architectures require lower color temperatures and additional optical effects to describe their history.

Based on the above observation and motivation, we propose a novel learnable context-aware 4D lookup table (4D~LUT), which enables content-dependent image enhancement without a significant increase in computational costs, achieving better visually pleasing results (as shown in Fig.~\ref{teaser}(c)). 
As shown in Fig.~\ref{teaser}(b), 4D~LUT extends the 3-dimensional lookup tables to a 4-dimensional space by introducing an additional contextual dimension, where the input index of 4D~LUT (\emph{i.e.,}~RGBC) varies with the context map, which increases the color enhancement capability of 4D~LUT and enabling finer control of color transformation and stronger image enhancement.
In particular, as shown in Fig.~\ref{Overview}, it includes four closely-related components. 1) We propose a context encoder to generate context maps through end-to-end learning. The context map represents the pixel-level category in an image based on their content difference, which can extend the image from RGB to RGBC. 
2) A parameter encoder for generating a group of coefficients through end-to-end learning. These coefficients can be adaptively changed according to the input image for assisting the final context-aware 4D~LUT generation.
3) Based on multiple pre-defined learnable basis 4D~LUTs and coefficients obtained above, we propose a 4D~LUTs fusion module that crosses different color spaces and integrates them into a final context-aware 4D~LUT with stronger enhanced capabilities.
4) We propose to use quadrilinear interpolation, which can convert the input image and context map into four-dimensional spatially indexed for input into the context-aware 4D~LUT, and finally outputs the enhanced image after the interpolation operation.
Compared with traditional 3D~LUT (\emph{i.e.,}~RGB mapping to RGB), the design of context-aware 4D~LUT (\emph{i.e.,}~RGBC mapping to RGB) encourages a content-dependent manner to enable finer control of color transformations, thereby enhancing the pixels' color from an input image to enhanced image.

Our contributions are summarized as follows:
\begin{itemize}
 \item To the best of our knowledge, we are the first to extend the lookup table architecture into a 4-dimensional space and achieve content-dependent image enhancement without a significant increase in computational costs. More specifically, we propose a learnable context-aware 4-dimensional lookup table (4D~LUT), which consists of four closely-related components context encoder, parameter encoder, 4D~LUTs fusion, and quadrilinear interpolation.
 \item The extensive experiments demonstrate that the proposed 4D~LUT can obtain more accurate results and significantly outperform existing SOTA methods in three widely-used image enhancement benchmarks.
\end{itemize}

The rest of the paper is organized as follows. Related work is reviewed in Sec.~\ref{Related Work}. The proposed context-aware 4D~LUT is elaborated in Sec.~\ref{Methodologyzz}. Experimental evaluation, analysis, and ablation study are presented in Sec.~\ref{Experimentszz}. The discussions of the related parameters and components are presented in Sec.~\ref{Discussions}.
Finally, we conclude this work in Sec.~\ref{Conclusion}.

\section{Related Work}
\label{Related Work}
In this section, we first briefly review the traditional algorithms for image enhancement and then review the recent popular study on deep learning-based algorithms.

\subsection{Traditional Algorithms}
Traditional image enhancement methods improve the visual quality of images by using hand-crafted global descriptors and local filters. For example, color correction~\cite{lu2015spatio} and color histogram equalization~\cite{xu2013generalized} adjust image colors by establishing color mapping relationships. Local laplacian filter~\cite{aubry2014fast} and Guided filter~\cite{he2012guided} enhance the image visual quality by operations such as detail smoothing/sharpening. However, only experienced experts can use these hand-crafted feature descriptors or filters, and they are costly in time.

\subsection{Learning-based Algorithms}
Recently, the learning-based image enhancement algorithm has been developed rapidly. These main algorithms can be categorized into three paradigms. Reinforcement learning-based methods, image-to-image translation methods, and physical modeling-based methods.

\subsubsection{Reinforcement learning-based methods}
The enhancement methods based on reinforcement learning~\cite{hu2018exposure,park2018distort,yu2018deepexposure,kosugi2020unpaired} attempt to decouple the whole process and gradually improve the image's visual quality by simulating the human step-by-step retouching process. Typically, White-Box~\cite{hu2018exposure} proposes to decouple the image enhancement process into a series of suitable parameters. And the deep reinforcement learning approach is used to learn the decision of what action to take next in the current state. Distort-and-Recover~\cite{park2018distort} casts a color enhancement process as a Markov Decision Process where actions are defined as global color adjustment operations. The agent is then trained to learn the optimal global enhancement sequence of the actions. Deepexposure~\cite{yu2018deepexposure} and UIE~\cite{kosugi2020unpaired} employ similar reinforcement learning strategies to learn an unpaired photo-enhanced model in an adversarial manner and severely sacrifice efficiency.

\subsubsection{Image-to-image translation methods}
The Image-to-image translation enhancement methods~\cite{isola2017image,zhu2017unpaired,chen2017photographic,liu2017unsupervised,chen2018deep,zamir2020learning,kim2020global,liu2022very} treat image enhancement as an image-to-image transformation problem, which tries to learn a mapping relationship between input and enhanced images through convolutional networks. 
Representatively, with the popularity of generative adversarial mechanisms~\cite{goodfellow2014generative}, some existing works~\cite{ignatov2017dslr,zhu2017unpaired,liu2017unsupervised} use the residual network~\cite{he2016deep} to style transfer and enhance unpaired images. Pix2Pix~\cite{isola2017image} investigates conditional adversarial networks as a general-purpose solution to image-to-image translation problems and learns a loss function to train the mapping from the input image to the output image. MIEGAN~\cite{pan2021miegan} presents a multi-module cascade generative network and an adaptive multi-scale discriminative network to capture both global and local information of a mobile image. DPE~\cite{chen2018deep} improves U-Net~\cite{ronneberger2015u} into a photo enhancer that transforms an input image into an enhanced image with the characteristics of given a set of photographs. GSGN~\cite{kneubuehler2020flexible} proposes the first practical multi-task image enhancement network, that is able to learn one-to-many and many-to-one image mappings. Focusing on the fact that subjectively people have diverse preferences for image aesthetics, PieNet~\cite{kim2020pienet} proposes the first deep learning method for personalized image enhancement that can enhance images for users by selecting preferences. MIRNet~\cite{zamir2020learning} proposes an architecture with the collective goals of maintaining spatially-precise high-resolution representations through the entire network and receiving strong contextual information from the low-resolution representations. CSRNet~\cite{he2020conditional,liu2022very} analyzes the mathematical formulation of image enhancement and proposes a lightweight framework consisting of $1\times 1$ convolutional layer.
However, these methods suffer from a lack of transparency in the whole enhancement process, like a ``black-box", which obscures their reliability.

\subsubsection{Physical modeling-based methods}
Inspired by the parametric (graduated, radial filters), brush tools, etc. in professional-grade software (\emph{e.g.,}~Photoshop, Lightroom), the enhancement based on physical model~\cite{guo2016lime,gharbi2017deep,wei2018deep,guo2020zero,wang2019underexposed,moran2020deeplpf,moran2021curl,zeng2020learning,li2021low} adjust the image color by predicting the relevant physics parameters and variables based on the proposed physical model and assumptions. Inspired by bilateral grid processing and local affine color transforms, HDRNet~\cite{gharbi2017deep} predicts the coefficients of a locally-affine model in bilateral space to approximate the desired image transformation. RetinexNet~\cite{wei2018deep,li2018structure} assumes that observed images can be decomposed into reflectance and illumination to enhance the low-light image. DeepUPE~\cite{wang2019underexposed} introduces intermediate illumination to associate the input with expected enhancement results for learning complex photographic adjustments. Zero-DCE~\cite{guo2020zero} formulates light enhancement as a task of image-specific curve estimation and estimates pixel-wise and high-order curves for dynamic range adjustment of an input image. SCI~\cite{ma2022toward} establishes a cascaded illumination learning process with weight sharing to handle the low-light enhancement task. DeepLPF~\cite{moran2020deeplpf} proposes learnable spatially local filters of three different types (Elliptical Filter, Graduated Filter, Polynomial Filter) and regresses the parameters of these filters that are then automatically applied to enhance the image. Inspired by the Photoshop curves tool, CURL~\cite{moran2021curl} designs a multi-color space neural retouching block and adjusts global image properties using human-interpretable image enhancement curves.

Especially, to improve the enhancement quality and efficiency, some existing works~\cite{liang2021ppr10k,zeng2020learning,wang2021real,yang2022adaint} propose to enhance images by improving the fundamental physical model 3D~LUT in digital image processing. These methods usually adopt a two-step solution, which includes 1) predicting the relevant coefficients based on the 3D~LUT to obtain an enhancer, and 2) adjusting the color based on the RGB value of each original pixel one by one to form an enhanced image through the uniform enhancer.
Although these methods spend less runtime on the enhancement process, they can only achieve a globally overall average over all regions of the enhanced result and lack the ability to finely control the color transformation of pixels with different content in each image. Therefore, in this paper, we extend the 3D~LUT into a 4D space and guide the content-dependent image enhancement by the additional dimensions (\emph{i.e.,}~context information) introduced.

\begin{figure*}
  \centering
  \includegraphics[width = 1.0\textwidth,page=2]{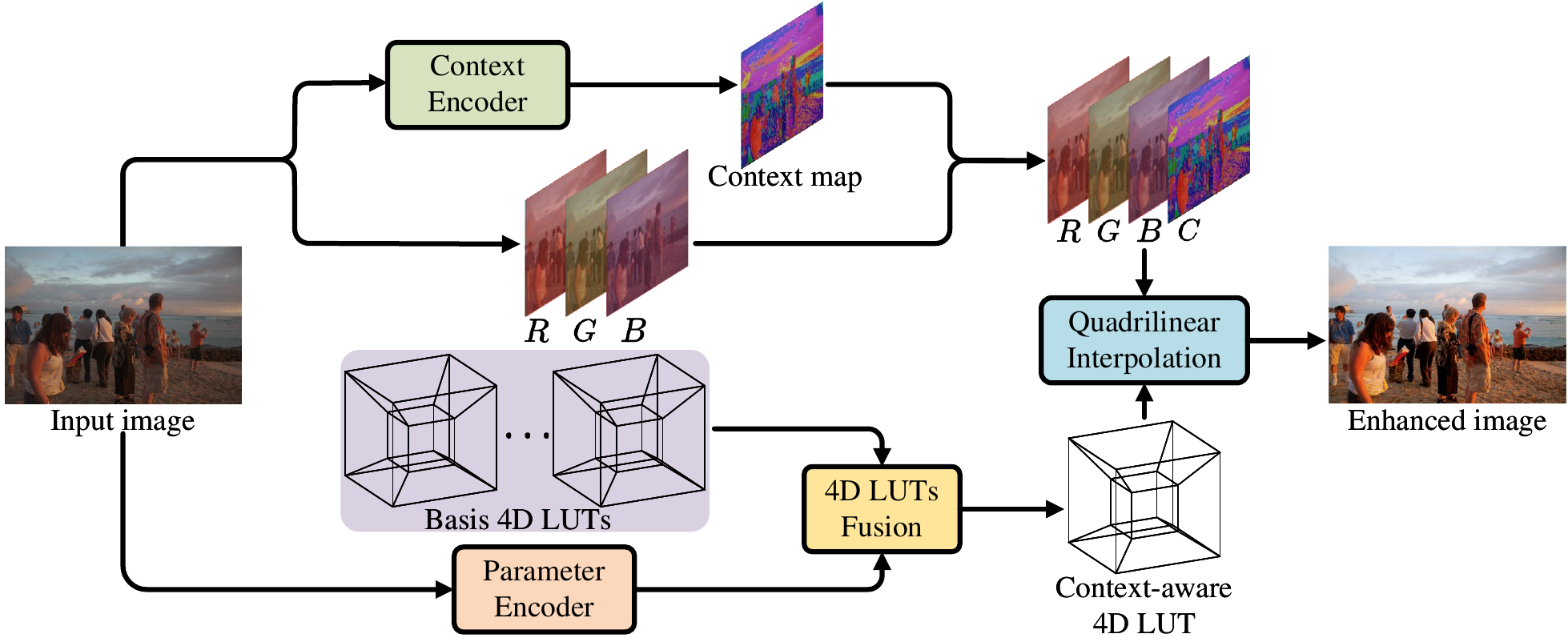}
  \caption{The overview of 4D~LUT. 4D~LUT takes an input image and pre-defined basis 4D~LUTs as input and generates an enhanced image. Context encoder for generating the pixel-level content-dependent context map. 
  Parameter encoder for generating image-adaptive coefficients. 4D~LUTs fusion module integrates the learnable basis 4D~LUTs and coefficients into a context-aware 4D~LUT. Quadrilinear interpolation module for generating the enhanced image by inputting them. Constrained by the optimization objective, the parameters of the entire network, including the context-aware 4D~LUT, can be trained end-to-end.}
  \label{Overview}
\end{figure*}

\section{Methodology}
\label{Methodologyzz}
In this section, to assist the understanding of the new proposed 4D~LUT, we first briefly review the preliminary about 3D~LUT and trilinear interpolation. Then we describe the overview and each component of 4D~LUT in detail.

\subsection{Preliminary}
\label{Preliminaryzz}

\subsubsection{3D~LUT}
\label{3D lookup tablezz}

3D~LUT is a common tool used in different camera imaging pipeline systems and software for image enhancement, which can be adjusted manually or algorithmically to achieve different kinds of enhancements. A 3D~LUT can be represented as a 3D lattice and the value of each element in the lattice can be represented as a triplet $(R_{out}^{(i,j,k)},G_{out}^{(i,j,k)},B_{out}^{(i,j,k)})$, where $i,j,k\in \{0,\dots,N_{bin}-1\}$, $N_{bin}$ is the number of bins along each of three dimensions. Such a lattice includes a total of $N_{bin}^{3}$ sampling points, which form a complete 3D color transformation space. As shown in Fig.~\ref{teaser}(a), the element $(r_{in},g_{in},b_{in})$ input to this color space can be mapped to an index by uniformly discretizing the RGB color space. The corresponding transformed output RGB color $(r_{out},g_{out},b_{out})$ is the value corresponding to the index. It is worth noting that as the value of $N_{bin}$ increases, the 3D color transformation space becomes more accurate for color transformation and vice versa.

\subsubsection{Trilinear interpolation}
\label{Trilinear interpolationzz}
As described above, the distribution of the elements in the LUT is discrete in space and it cannot be sampled directly by the input index. Therefore, when sampling elements in the 3D~LUT, the input color will find its nearest sample point based on its index and calculate its transformed output by trilinear interpolation~\cite{liang2021ppr10k,zeng2020learning}.

Specifically, to locate the nearest 8 adjacent elements around input index, we first construct the input index $(x,y,z)$ to the 3D~LUT based on the input RGB color $(r_{in}^{(x,y,z)},g_{in}^{(x,y,z)},b_{in}^{(x,y,z)})$, which process can be described as follows:
\begin{equation}
    x=r_{in}^{(x,y,z)}\cdot \frac{N_{bin}}{255},y=g_{in}^{(x,y,z)}\cdot \frac{N_{bin}}{255},z=b_{in}^{(x,y,z)}\cdot \frac{N_{bin}}{255},
\end{equation}
where $N_{bin}$ is the number of bins along each of three dimensions in 3D~LUT. We use $(i,j,k)$ to denote the location of the defined sampling point, which can be calculated as follows:
\begin{equation}
    i=\left\lfloor x\right\rfloor,j=\left\lfloor y\right\rfloor,k=\left\lfloor z\right\rfloor,
\end{equation}
where $\left\lfloor \cdot \right\rfloor$ represents the floor function. We use $(o_{x},o_{y},o_{z})$ denote the offset of the input index $(x,y,z)$ to defined sampling point $(i,j,k)$, which can be calculated as follows:
\begin{equation}
   o_{x}=x-i,o_{y}=y-j,o_{z}=z-k.
\end{equation}
Then, taking the red color $r_{out}^{(x,y,z)}$ in transformed output RGB color $(r_{out}^{(x,y,z)},g_{out}^{(x,y,z)},b_{out}^{(x,y,z)})$ as an example, the interpolation process can be expressed as:
\begin{equation}
\scriptsize
\begin{aligned}
    r_{out}^{(x,y,z)}&=(1-o_x)(1-o_y)(1-o_z)r_{out}^{(i,j,k)}+o_xo_yo_zr_{out}^{(i+1,j+1,k+1)}\\
    &+o_x(1-o_y)(1-o_z)r_{out}^{(i+1,j,k)}+(1-o_x)o_yo_zr_{out}^{(i,j+1,k+1)}\\
    &+(1-o_x)o_y(1-o_z)r_{out}^{(i,j+1,k)}+o_x(1-o_y)o_zr_{out}^{(i+1,j,k+1)}\\
    &+(1-o_x)(1-o_y)o_zr_{out}^{(i,j,k+1)}+o_xo_y(1-o_z)r_{out}^{(i+1,j+1,k)}\\
\end{aligned}
\end{equation}
where the input $r_{out}^{(\{i,i+1\},\{j,j+1\},\{k,k+1\})}$ is the transformed output red color corresponding to the defined sampling point $(\{i,i+1\},\{j,j+1\},\{k,k+1\})$. Similarly, it can also obtain the other colors (\emph{i.e.,}~$g_{out}^{(x,y,z)}$ and $b_{out}^{(x,y,z)}$) in the same way. It is worth noting that the interpolation operation is differentiable and can update of LUTs during end-to-end training.

\subsection{Context-aware 4D~LUT}
Existing works~\cite{liang2021ppr10k,zeng2020learning,wang2021real,yang2022adaint} improve the 3D~LUT to enhance the image, and lack a necessary content information in the images. Therefore, we propose the learnable context-aware 4D~LUT to achieve content-dependent image enhancement and enable finer control of color transformations for pixels with different content in each image.

As shown in Fig.~\ref{Overview}, our method takes the input image and several pre-defined basis 4D~LUTs as input, and finally generates an enhanced image. 
It includes four closely-related components, context encoder, parameter encoder, 4D~LUTs fusion module, and quadrilinear interpolation module. 
Specifically, it involves the following stages: 
1) We first use a context encoder to generate a context map that represents the pixel-level category from the input image through end-to-end learning. 
2) Parallelly, we use a parameter encoder for generating image-adaptive coefficients applied to fuse the learnable pre-defined basis 4D~LUTs. 
3) Then, based on the output of the parameter encoder, we use the 4D~LUTs fusion module to integrate the learnable basis 4D~LUTs into a final context-aware 4D~LUT with more enhanced capabilities. 
4) Finally, the original input image and the context-aware 4D~LUT are input to the quadrilinear interpolation module to obtain the enhanced image. 
In the following, we will describe them individually.

\subsubsection{Context Encoder}
\label{Context Encoderzz}
Context encoder can adaptively generate content-dependent context maps under the constraints of the objective function in a learnable manner. We use $E_{context}(\cdot)$ to denote the context encoder, which consists of a series of stacked residual blocks. 
In detail, it includes four residual blocks with $3 \times3$ kernel size and one residual block with $1 \times1$ size. 
Among them, the $3 \times3$ residual blocks are applied to extract the high-level image features of the same resolution from the input image, and the $1 \times1$ residual block compresses the image features and is used to output the context map. 
Suppose that an input image $I_{input}\in R^{3\times H\times W}$ are given. The generated context map $C\in R^{1\times H\times W}$ can be formulated as:
\begin{equation}
   C=E_{context}(I_{input}).
\end{equation}

Intuitively, the context encoder can be viewed as a function that maps the content information for each position in the input image to a compact scalar representation. During training, the context encoder is updated by the back-propagated gradient of the module connected behind. During inference, the more appropriate context map for enhancement is generated adaptively according to the high-level semantic differences of different regions.

\subsubsection{Parameter Encoder}
To facilitate the fusion of multiple pre-defined basis 4D~LUTs and increase the 4D~LUT enhancement capability, the parameter encoder extracts a group of image-adaptive coefficients for 4D~LUT fusion during end-to-end training. We use $E_{param}(\cdot)$ to denote the parameter encoder, which consists of a series of stacked residual blocks and a parameters output layer consisting of a convolutional layer. 
In detail, suppose that an input image $I_{input}\in R^{3\times H\times W}$ are given. The generated parameters $\mathcal{W}\in R^{N_{w}\times 1\times 1}$ and $\mathcal{B}\in R^{N_{b}\times 1\times 1}$ can be formulated as:
\begin{equation}
   \mathcal{W},\mathcal{B}=E_{param}(I_{input}),
\end{equation}
where $\mathcal{W}=\{w_1,w_2,\dots,w_{N_w}\}$ and $\mathcal{B}=\{b_1,b_2,\dots,b_{N_b}\}$ represent the outputted coefficients (\emph{i.e.,}~weights and biases) to fuse the learnable basis 4D~LUTs, respectively. $N_w$ and $N_b$ are the number of weights and biases in the coefficients, respectively. If the number of basis 4D~LUTs is assumed to be $N_{lut}$, then the values of $N_w$ and $N_b$ are $3N_{lut}^2$ and $N_{lut}$, respectively.

During training, the parameter encoder is updated by the back-propagated gradient of the module connected behind. During inference, the parameter encoder can be viewed as a parameter predictor that integrates the learnable basis 4D~LUTs in a soft-weighting strategy to achieve adaptive context-aware 4D~LUT generation for better image enhancement.

\subsubsection{4D~LUTs Fusion}
\label{4DLUTs Fusionzz}
During the end-to-end training process, the elements of per-defined basis 4D~LUTs are gradually updated to adapt to the change of color space. To obtain a context-aware 4D~LUT with stronger color transformation capabilities, the 4D~LUTs fusion module fuses multiple learnable basis 4D~LUTs by using the coefficients obtained from the parameter encoder.

Specifically, we use $\Psi_{n\in\{1,\dots,N_{lut}\}}$ to denote one of the multiple basis 4D~LUTs, where $N_{lut}$ is the number of basis 4D~LUTs. As described in Sec.~\ref{3D lookup tablezz}, the value of each element in $\Psi_{n}$ can be represented as a triplet (\emph{i.e.,}~red, green, and blue color spaces), in which we use $\psi^{r}_n$, $\psi^{g}_n$, and $\psi^{b}_n$ to denote the corresponding red, green, and blue color spaces in $\Psi_{n}$, respectively. Besides, we use $\hat{\Psi}=(\hat{\psi}^{r},\hat{\psi}^{g},\hat{\psi}^{b})$ to represent the fused context-aware 4D~LUT. Taking the fusion process of red space $\hat{\psi}^{r}$ as an example, it can be generated by:
\begin{equation}
\begin{aligned}
      \hat{\psi}^{r} &= \textstyle \sum_{n=1}^{N_{lut}}(w_{n}\psi^{r}_n+w_{(N_{lut}+n)}\psi^{g}_n \\
      &+w_{(2N_{lut}+n)}\psi^{b}_n+b_{n}),
\end{aligned}
\end{equation}
where $w$ and $b$ are the weights and biases output from the parameter encoder. Similarly, we can also obtain the other colors space(\emph{i.e.,}~$\hat{\psi}^{g}$ and $\hat{\psi}^{b}$) in the same way. 

In general, the addition of weights allows the different color spaces to interact and fuse with each other, resulting in a more appropriate color temperature (similar to white balance). The addition of biases can adaptively enhance the overall brightness of the image. Such a design on fusion approach makes the fused context-aware 4D~LUT with more superior enhancement capability.

\subsubsection{Quadrilinear Interpolation}
Based on the original RGB image, the generated context map, and context-aware 4D~LUT, we can obtain the enhanced image via interpolation operation. However, different from the 3D~LUT and trilinear interpolation described in Sec.~\ref{Preliminaryzz}, the index of our proposed 4D~LUT is on the 4-dimensional space (\emph{i.e.,}~RGB+Context). 

To effectively interpolate the values based on the index of 4-dimensional RGBC space, we propose to use a quadrilinear interpolation closely related to the 4D~LUT. Suppose that an input image $I_{input}\in R^{3\times H\times W}$ are given, the context map $C\in R^{1\times H\times W}$ and context-aware 4D~LUT are generated, the enhanced image $I_{output}\in R^{3\times H\times W}$ can be formulated as:
\begin{equation}
   I_{output}=QI_{\hat{\Psi}}(Concat(I_{input},C)),
\end{equation}
where $QI_{\hat{\Psi}}(\cdot)$ denotes the quadrilinear interpolation based on the context-aware 4D~LUT $\hat{\Psi}$. $Concat(\cdot)$ is the concatenation operation.

Specifically, similar to in Sec.~\ref{Trilinear interpolationzz} above, the input index to the 4D~LUT based on the input RGBC value can be represented as $(x,y,z,u)$, we first locate the nearest 16 adjacent elements around the input index as the sampling point (\emph{i.e.,}~the distance from input index satisfies $\Delta \in [-1,1]$ in each dimension).
We use $(i,j,k,l)$ to denote the coordinates of a defined sampling point in 4D~LUT, which can be calculated as follows:
\begin{equation}
    i=\left\lfloor x\right\rfloor,j=\left\lfloor y\right\rfloor,k=\left\lfloor z\right\rfloor,l=\left\lfloor u\right\rfloor,
\end{equation}
where $\left\lfloor \cdot \right\rfloor$ represents the floor function. The defined nearest 16 adjaent sampling point in 4D~LUT are $(\{i,i+1\},\{j,j+1\},\{k,k+1\},\{l,l+1\})$.
We use $(o_{x},o_{y},o_{z},o_{u})$ denote the offset of the input index $(x,y,z,u)$ to defined sampling point $(i,j,k,l)$, which can be calculated as follows:
\begin{equation}
   o_{x}=x-i,o_{y}=y-j,o_{z}=z-k,o_{u}=u-l.
\end{equation}
Then, the red color $r_{out}^{(x,y,z,u)}$ in transformed output RGB color $(r_{out}^{(x,y,z,u)},g_{out}^{(x,y,z,u)},b_{out}^{(x,y,z,u)})$ can be expressed as:
\begin{equation}
\scriptsize
\begin{aligned}
    r_{out}^{(x,y,z,u)}&=(1-o_x)(1-o_y)(1-o_z)(1-o_u)r_{out}^{(i,j,k,l)} \\
    &=o_x(1-o_y)(1-o_z)(1-o_u)r_{out}^{(i+1,j,k,l)}\\
    &=(1-o_x)o_y(1-o_z)(1-o_u)r_{out}^{(i,j+1,k,l)}\\
    &=(1-o_x)(1-o_y)o_z(1-o_u)r_{out}^{(i,j,k+1,l)}\\
    &=(1-o_x)(1-o_y)(1-o_z)o_ur_{out}^{(i,j,k,l+1)}\\
    &=o_xo_y(1-o_z)(1-o_u)r_{out}^{(i+1,j+1,k,l)}\\
    &=(1-o_x)o_yo_z(1-o_u)r_{out}^{(i,j+1,k+1,l)}\\
    &=(1-o_x)(1-o_y)o_zo_ur_{out}^{(i,j,k+1,l+1)}\\
    &=o_x(1-o_y)o_z(1-o_u)r_{out}^{(i+1,j,k+1,l)}\\
    &=o_x(1-o_y)(1-o_z)o_ur_{out}^{(i+1,j,k,l+1)}\\
    &=(1-o_x)o_y(1-o_z)o_ur_{out}^{(i,j+1,k,l+1)}\\
    &=o_xo_yo_z(1-o_u)r_{out}^{(i+1,j+1,k+1,l)}\\
    &=(1-o_x)o_yo_zo_ur_{out}^{(i,j+1,k+1,l+1)}\\
    &=o_xo_y(1-o_z)o_ur_{out}^{(i+1,j+1,k,l+1)}\\
    &=o_x(1-o_y)o_zo_ur_{out}^{(i+1,j,k+1,l+1)}\\
    &+o_xo_yo_zo_ur_{out}^{(i+1,j+1,k+1,l+1)},
\end{aligned}
\end{equation}
where the input $r_{out}^{(\{i,i+1\},\{j,j+1\},\{k,k+1\},\{l,l+1\})}$ is the transformed output red color corresponding to the defined nearest 16 adjacent sampling point $(\{i,i+1\},\{j,j+1\},\{k,k+1\},\{l,l+1\})$ in 4D~LUT. Similarly, we can also obtain the other colors (\emph{i.e.,}~$g_{out}^{(x,y,z)}$ and $b_{out}^{(x,y,z)}$) in the same way. The quadrilinear interpolation operation is differentiable and can propagate the gradient to update the weight of the network and the element values of context-aware 4D~LUT.

\subsection{Training}
To update the elements of 4D~LUT and the parameters of the network, in this section, we employ several objective functions to supervise the whole training process.

\subsubsection{4D smooth regularization}
\label{4D smooth regularization}
To avoid artifacts caused by extreme color changes in the 4D~LUT, it is necessary to ensure that the converting from the input space (\emph{i.e.,}~RGBC) to the obtained color space (\emph{i.e.,}~RGB) is stable enough. We introduce $L_{2}$-norm regularization on the elements of the 4D~LUT and the outputted coefficient of the parameter encoder to improve the smoothness of the context-aware 4D~LUT.

Specifically, inspired by existing work~\cite{zeng2020learning}, we extend the 3D smooth regularization into a 4D smooth regularization term on the learning of 4D~LUT to ensure the local smoothing of the elements in 4D~LUT. The smooth regularization of 4D~LUT $L_{s}^{lut}$ can be calculated as follow:
\begin{equation}
\begin{aligned}
    L_{s}^{lut}&=\sum_{p\in\{r,g,b\}}\sum_{i,j,k,l=0}^{N_{bin}-1}(\big \|p^{(i+1,j,k,l)}_{out}-p^{(i,j,k,l)}_{out}\big \|^{2} \\
    &+\big \|p^{(i,j+1,k,l)}_{out}-p^{(i,j,k,l)}_{out} \big \|^{2}+\big \|p^{(i,j,k+1,l)}_{out} \\
    &-p^{(i,j,k,l)}_{out}\big \|^{2}+\big \|p^{(i,j,k,l+1)}_{out}-p^{(i,j,k,l)}_{out}\big \|^{2}),
\end{aligned}
\end{equation}
where $N_{bin}$ is the number of bins along each of the output dimensions in 4D~LUT. The input $p_{out}^{(\{i,i+1\},\{j,j+1\},\{k,k+1\},\{l,l+1\}))}$ is the transformed output red, green, and blue color corresponding to the defined sampling point $(\{i,i+1\},\{j,j+1\},\{k,k+1\},\{l,l+1\})$ in 4D~LUT.

The smooth regularization of outputted coefficient $L_{s}^{coe}$ can be calculated as follow:
\begin{equation}
    L_{s}^{coe}=\sum_{n=1}^{N_w}\big \|w_n\big \|^{2}+\sum_{m=1}^{N_b}\big \|b_n\big \|^{2},
\end{equation}
where $N_w$ and $N_b$ are the numbers of weights and biases in the coefficients, respectively. $w_n$ and $b_m$ represent the outputted image-adaptive weights and biases from the parameter encoder.

The overall smooth regularization term $L_{s}$ can be represented as:
\begin{equation}
\label{1}
    L_{s}=L_{s}^{lut}+L_{s}^{coe}.
\end{equation}
This design makes the elements in 4D~LUT locally smoother and ensures the stability of color transformation.

\subsubsection{4D monotonicity regularization}
\label{4D monotonicity regularization}
To preserve the robustness and relative color brightness/saturation in the enhancement process, we followed existing work~\cite{zeng2020learning} to expand the 3D monotonicity regularization into 4D monotonicity regularization term $L_{m}$ as follows:
\begin{equation}
\label{2}
\begin{aligned}
    L_{m}&=\sum_{p\in\{r,g,b\}}\sum_{i,j,k,l=0}^{N_{bin}-1}[g(p^{(i,j,k,l)}_{out}-p^{(i+1,j,k,l)}_{out}) \\
    &+g(p^{(i,j,k,l)}_{out}-p^{(i,j+1,k,l)}_{out})+g(p^{(i,j,k,l)}_{out} \\
    &-p^{(i,j,k+1,l)}_{out})+g(p^{(i,j,k,l)}_{out}-p^{(i,j,k,l+1)}_{out})],
\end{aligned}
\end{equation}
where $g(\cdot)$ denotes the ReLU activation function (\emph{i.e.,}~$g(x)=max(0,x)$). $N_{bin}$ is the number of bins along each of the output dimensions in 4D~LUT. The input $p_{out}^{(\{i,i+1\},\{j,j+1\},\{k,k+1\},\{l,l+1\}))}$ is the transformed output red, green, and blue color corresponding to the defined sampling point $(\{i,i+1\},\{j,j+1\},\{k,k+1\},\{l,l+1\})$ in 4D~LUT. This design not only allows the color transformation to cover the entire RGBC space, but also allows the 4D~LUT to converge faster during training.

\subsubsection{Pairwise reconstruction}
For fair comparisons, we follow previous works~\cite{gharbi2017deep,zeng2020learning,liu2022very} to define the reconstruction loss $L_{r}$ between the ground truth $I_{GT}$ and enhanced image $I_{output}$ to train the whole model, it is defined as:
\begin{equation}
L_{r} = \frac{1}{N_{bs}}\sum^{N_{bs}}_{1} (I_{GT}-I_{output})^2,
\end{equation}
where $N_{bs}$ represents the batch size during training.

\subsubsection{Loss function}
As described above, during training, our approach includes a total of three loss components, the 4D smooth regularization loss, the 4D monotonicity regularization loss, and the pairwise reconstruction loss. It is essential for the network to balance these three items. Therefore, we multiply $L_{s}$ and $L_{m}$ (as described in Equ.~\ref{1} and~\ref{2}) with a weight $\alpha_s$ and $\alpha_{m}$, respectively, to enable the validity of context-aware 4D~LUT while not harming the performance of enhancement. The total loss function is formulated as follows: 
\begin{equation} 
\label{lossfunctionzz}
L_{total}=L_{r}+\alpha_sL_{s}+\alpha_mL_{m}.
\end{equation}

\section{Experiments}
\label{Experimentszz}

\subsection{Experimental Settings}

\subsubsection{Datasets}
We evaluate the proposed 4D~LUT and compare its performance with other state-of-the-art (SOTA) approaches on three widely-used challenging benchmarks, derived from two public datasets: \textbf{MIT-Adobe-5K-UPE}~\cite{wang2019underexposed}, \textbf{MIT-Adobe-5K-DPE}~\cite{chen2018deep}, and \textbf{PPR10K}~\cite{liang2021ppr10k}.

\textbf{MIT-Adobe-5K-UPE:} a benchmark is divided from the  MIT-Adobe FiveK dataset~\cite{bychkovsky2011learning}, following the dataset pre-processing procedure of DeepUPE~\cite{wang2019underexposed}. MIT-Adobe FiveK dataset is a commonly-used landscape photo retouching dataset with 5,000 images captured using various DSLR cameras. Each image contains the corresponding retouched version produced by five experienced experts (A/B/C/D/E). For fair comparisons, we follow previous works~\cite{chen2018deep,moran2020deeplpf,moran2021curl} to use the photo retouched by expert C as image enhancement ground truth (GT). We select 4,500 image pairs as the training set and 500 image pairs as the test set in order, and all images are resized to 510 pixels on the long edge.

\textbf{MIT-Adobe-5K-DPE:} another benchmark consists of the same image context as the MIT-Adobe FiveK dataset~\cite{bychkovsky2011learning} however following the dataset pre-processing procedure of DPE~\cite{chen2018deep}. We follow previous works~\cite{wang2019underexposed,moran2020deeplpf,moran2021curl} to use the photo retouched by expert C as ground truth (GT). The difference is that we sequentially select 2,250 image pairs as the training set and 500 image pairs as the test set.

\textbf{PPR10K:} a new large-scale portrait retouching dataset to be released in 2021, containing a total of 11,161 high-quality RAW portraits. Each image contains the corresponding retouched version produced by three experienced experts (a/b/c). For fair comparisons, we follow the official split~\cite{liang2021ppr10k} to divide the dataset into 8,875 training pairs and 2,286 test pairs. We compare the results on all expert modifications, and the image size is 360p.

\subsubsection{Evaluation metrics}
For fair comparisons, we follow previous enhancement works~\cite{gharbi2017deep,zeng2020learning,liu2022very} to use peak signal-to-noise ratio (PSNR) and structural similarity index (SSIM)~\cite{wang2004image} as a commonly-used metric for evaluating in terms of the color and structure similarity between the enhanced results and the corresponding expert-retouched images. 

\subsubsection{Implementation details}
Our experiment is conducted on an NVIDIA 2080Ti GPU through PyTorch. For fair comparisons, we follow previous works~\cite{liang2021ppr10k,zeng2020learning,wang2021real,yang2022adaint} and use the Adam optimizer~\cite{kingma2014adam} with $\beta_{1}=0.9$ and $\beta_{2}=0.999$ and the batch size of $1$. The initial learning rate is set as $1 \times 10^{-4}$ and then reduce the learning rate by a factor of $0.2$ when the losses of the testing set last for $20$ epochs without decreasing. We jointly train the entire model for $400$ epochs. Except for adding random crop image patches with a scale in the range $[0.6,1.0]$, horizontal flip, and no other data augmentation methods are used. 

Besides, we follow previous works~\cite{liang2021ppr10k,zeng2020learning} to set the the number of bins $N_{bin}$ in LUT and the number of basis 4D~LUT $N_{lut}$ as 33 and 3, respectively. We set the number of weights $N_w$ and biases $N_b$ as 27 and 3, respectively. We empirically set $\alpha_s$ and $\alpha_m$ as 0.0001 and 10 through discussion experiments.

\begin{table}[!t]
\begin{center}
\caption{Quantitative comparison (PSNR$\uparrow$ and SSIM$\uparrow$) on the MIT-Adobe-5K-UPE~\cite{wang2019underexposed} dataset. \textcolor{red}{Red} indicates the best and \underline{\textcolor{blue}{blue}} indicates the second best performance (best view in color).}
\label{table:1}
\begin{tabular}{l  | c  | c  }
\hline

\hline
Method  & PSNR(dB)$\uparrow$ & SSIM$\uparrow$ \\
\hline
HDRNet~\cite{gharbi2017deep}                              & 21.96 & 0.866  \\
U-Net~\cite{ronneberger2015u}                            & 22.24 & 0.850 \\
DPE~\cite{chen2018deep}                                   & 22.15 & 0.850\\
White-Box~\cite{hu2018exposure}                           & 18.57 & 0.701\\
Dis-Rec~\cite{park2018distort}                           & 20.97 & 0.841 \\
DeepUPE~\cite{wang2019underexposed}                      & 23.04 & 0.893 \\
MIRNet~\cite{zamir2020learning}                            & 23.73 & 0.897 \\
TED+CURL~\cite{moran2021curl}                              & 24.20 & 0.880 \\
DeepLPF~\cite{moran2020deeplpf}                             &24.48 & 0.887 \\
CSRNet~\cite{he2020conditional,liu2022very}                 & 24.23 & 0.900 \\
3D~LUT~\cite{zeng2020learning}                              & \underline{\textcolor{blue}{24.60}} & \underline{\textcolor{blue}{0.911}} \\
\hline
\textbf{4D~LUT(Ours)}      & \textcolor{red}{24.96} & \textcolor{red}{0.924} \\
\hline

\hline
\end{tabular}
\end{center}
\end{table}

\begin{table}[!t]
\begin{center}
\caption{Quantitative comparison (PSNR$\uparrow$ and SSIM$\uparrow$) on the MIT-Adobe-5K-DPE~\cite{chen2018deep} dataset. \textcolor{red}{Red} indicates the best and \underline{\textcolor{blue}{blue}} indicates the second best performance (best view in color).}
\label{table:2}
\begin{tabular}{l  | c | c  }
\hline

\hline
Method  & PSNR(dB)$\uparrow$ & SSIM$\uparrow$ \\
\hline
DPED~\cite{ignatov2017dslr}                              & 21.76 & 0.871  \\
8Resblock~\cite{zhu2017unpaired,liu2017unsupervised}     & 23.42 & 0.875\\
CRN~\cite{chen2017photographic}                          & 22.38 & 0.877\\
U-Net~\cite{ronneberger2015u}                            & 22.13 & 0.879 \\
White-Box~\cite{hu2018exposure}                           & 21.32 & 0.864\\
Dis-Rec~\cite{park2018distort}                           & 21.60 & 0.875 \\
DPE~\cite{chen2018deep}                                   & 23.80 & 0.900\\
UIE~\cite{kosugi2020unpaired}                             & 22.27 & 0.881 \\
DeepLPF~\cite{moran2020deeplpf}                             & 23.93 & 0.903 \\
TED+CURL~\cite{moran2021curl}                              & 24.08 & 0.900 \\
GSGN~\cite{kneubuehler2020flexible}                         & 24.16 & 0.905 \\
3D~LUT~\cite{zeng2020learning}                              & \underline{\textcolor{blue}{24.33}} & \underline{\textcolor{blue}{0.910}} \\
\hline
\textbf{4D~LUT(Ours)}      & \textcolor{red}{24.61} & \textcolor{red}{0.918} \\
\hline

\hline
\end{tabular}
\end{center}
\end{table}

\subsection{Comparison with State-of-the-art Methods}
We compare our 4D~LUT with other classical start-of-the-art methods. 
These methods can be summarized into three categories: reinforcement learning-based methods (\emph{i.e.,}~White-Box~\cite{hu2018exposure}, Dis-Rec~\cite{park2018distort}, and UIE~\cite{kosugi2020unpaired}), image-to-image translation methods (\emph{i.e.,}~DPED~\cite{ignatov2017dslr}, 8Resblock~\cite{zhu2017unpaired,liu2017unsupervised}, CRN~\cite{chen2017photographic}, U-Net~\cite{ronneberger2015u}, DPE~\cite{chen2018deep}, GSGN~\cite{kneubuehler2020flexible}, MIRNet~\cite{zamir2020learning}, and CSRNet~\cite{he2020conditional,liu2022very}), and physical modeling-based methods (\emph{i.e.,}~HDRNet~\cite{gharbi2017deep}, DeepUPE~\cite{wang2019underexposed}, DeepLPF~\cite{moran2020deeplpf}, TED+CURL~\cite{moran2021curl}, 3D LUT~\cite{zeng2020learning}, 3D~LUT+HRP~\cite{liang2021ppr10k}).
For fair comparisons, we obtain the performance from their original paper or reproduce results with recommended configurations by the authors' officially released models.

\begin{table*}[!t]
\begin{center}
\caption{Quantitative comparison (PSNR$\uparrow$ and SSIM$\uparrow$) on the PPR10K~\cite{liang2021ppr10k} dataset. \textcolor{red}{Red} indicates the best and \underline{\textcolor{blue}{blue}} indicates the second best performance (best view in color).}
\label{table:3}
\begin{tabular}{l | c | c | c | c | c | c | c | c  }
\hline

\hline
\multirow{2}*{Method}  &  \multirow{2}*{Runtime} &  \multirow{2}*{\#Param} & \multicolumn{2}{c|}{PPR10K-a} & \multicolumn{2}{c|}{PPR10K-b} & \multicolumn{2}{c}{PPR10K-c} \\
\cline{4-9} 
~ &  ~ &  ~ & PSNR(dB)$\uparrow$ & SSIM$\uparrow$ & PSNR(dB)$\uparrow$ & SSIM$\uparrow$ & PSNR(dB)$\uparrow$ & SSIM$\uparrow$ \\
\hline
HDRNet~\cite{gharbi2017deep}                     & 6.03ms & 482K   & 21.435 & 0.905 & 21.609 & 0.907 & 21.841 & 0.903  \\
DeepLPF~\cite{moran2020deeplpf}                  & 51.3ms & 1.7M & 23.961 & 0.930 & 22.556 & 0.919 & 22.763 & 0.907\\
TED+CURL~\cite{moran2021curl}                    & 82.3ms & 1.4M & 23.651 & 0.914 & 23.324 & 0.911 & 23.869 & 0.903\\
CSRNet~\cite{he2020conditional,liu2022very}       & 2.50ms & 36.4K  & 24.039 & 0.935 & 24.066 & 0.939 & 24.257 & 0.930\\
3D LUT~\cite{zeng2020learning}           & 1.99ms & 593.5K  & \underline{\textcolor{blue}{24.632}} & 0.937 & \underline{\textcolor{blue}{24.101}} & 0.937  & \underline{\textcolor{blue}{24.515}} & 0.924\\
3D~LUT+HRP~\cite{liang2021ppr10k}    & 1.99ms & 593.5K & 24.416 & \underline{\textcolor{blue}{0.942}} & 23.985 & \underline{\textcolor{blue}{0.941}}   & 24.291 & \underline{\textcolor{blue}{0.930}}\\
\hline
\textbf{4D~LUT(Ours)}   & 5.75ms    &  924.4K     & \textcolor{red}{24.915} & \textcolor{red}{0.944}  & \textcolor{red}{24.398} & \textcolor{red}{0.942}  & \textcolor{red}{24.733} & \textcolor{red}{0.933}     \\
\hline

\hline
\end{tabular}
\end{center}
\end{table*}

\begin{figure*}[t!]
  \centering
  \includegraphics[width = 1.0\textwidth,page=3]{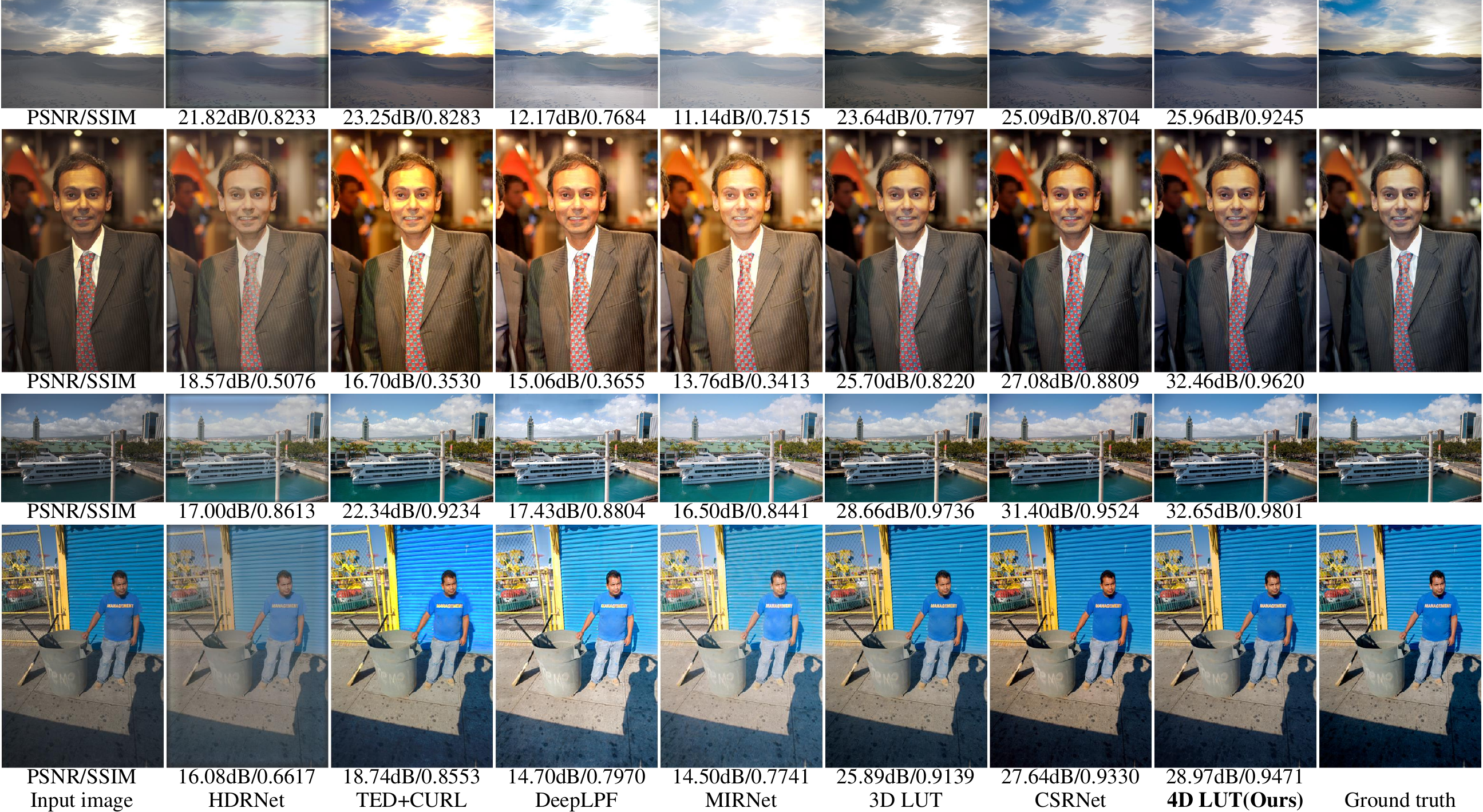}
  \caption{Visual comparison with state-of-the-arts on MIT-Adobe-5K-UPE~\cite{wang2019underexposed} dataset. The quantitative comparison (PSNR$\uparrow$ and SSIM$\uparrow$) is shown at the bottom of each case.}
  \label{fig:case_UPE}
\end{figure*}

\subsubsection{Quantitative comparison}
The results of each algorithm evaluated on datasets MIT-Adobe-5K-UPE~\cite{wang2019underexposed} and MIT-Adobe-5K-DPE~\cite{chen2018deep} are shown in Tab.~\ref{table:1} and Tab.~\ref{table:2}, respectively.
Benefit from the strong capabilities of a pure CNN structure based on an image-to-image translation methods (\emph{i.e.,}~DPED~\cite{ignatov2017dslr}, 8Resblock~\cite{zhu2017unpaired,liu2017unsupervised}, CRN~\cite{chen2017photographic}, U-Net~\cite{ronneberger2015u}, DPE~\cite{chen2018deep}, GSGN~\cite{kneubuehler2020flexible}, MIRNet~\cite{zamir2020learning}, and CSRNet~\cite{he2020conditional,liu2022very}) perform image enhancement by designing huge and complex network models, whose performance is often positively correlated with the model size. The latest algorithm CSRNet~\cite{he2020conditional,liu2022very} designs a lightweight enhancement model using the $1\times 1$ convolutional kernels, but still lacks enhancement capability. Besides, algorithms based on reinforcement learning (\emph{i.e.,}~White-Box~\cite{hu2018exposure}, Dis-Rec~\cite{park2018distort}, and UIE~\cite{kosugi2020unpaired}) improve the enhancement capability by decoupling multiple steps and also achieve pleasant results, but the computational cost is too large. The physical model-based methods (\emph{i.e.,}~HDRNet~\cite{gharbi2017deep}, DeepUPE~\cite{wang2019underexposed}, DeepLPF~\cite{moran2020deeplpf}, TED+CURL~\cite{moran2021curl}, 3D~LUT~\cite{zeng2020learning}, 3D~LUT+HRP~\cite{liang2021ppr10k}) is based on theoretical physical models and assumptions that are transparent in the enhancement process. Representatively, the latest algorithm 3D~LUT~\cite{zeng2020learning} generally performs better than the other methods. However, this method focus on learning a uniform enhancer and achieving a globally overall average over all regions of the enhanced result that decrease the accuracy of enhancement and can lead to sub-optimal performance. 

Our proposed 4D~LUT extends the lookup table architecture into a 4-dimensional space and achieves content-dependent image enhancement. As shown in Tab.~\ref{table:1} and Tab.~\ref{table:2}, it achieves a result of 24.96dB and 24.61dB PSNR and significantly outperforms the other algorithms for all datasets by a large margin. Specifically, on the MIT-Adobe-5K-UPE~\cite{wang2019underexposed} and MIT-Adobe-5K-DPE~\cite{chen2018deep} datasets, 4D~LUT outperforms 3D~LUT~\cite{zeng2020learning} by \textbf{0.36dB} and \textbf{0.28dB}, respectively. This large margin demonstrates the power of 4D~LUT in image enhancement.

\begin{figure*}[t!]
  \centering
  \includegraphics[width = 1.0\textwidth,page=9]{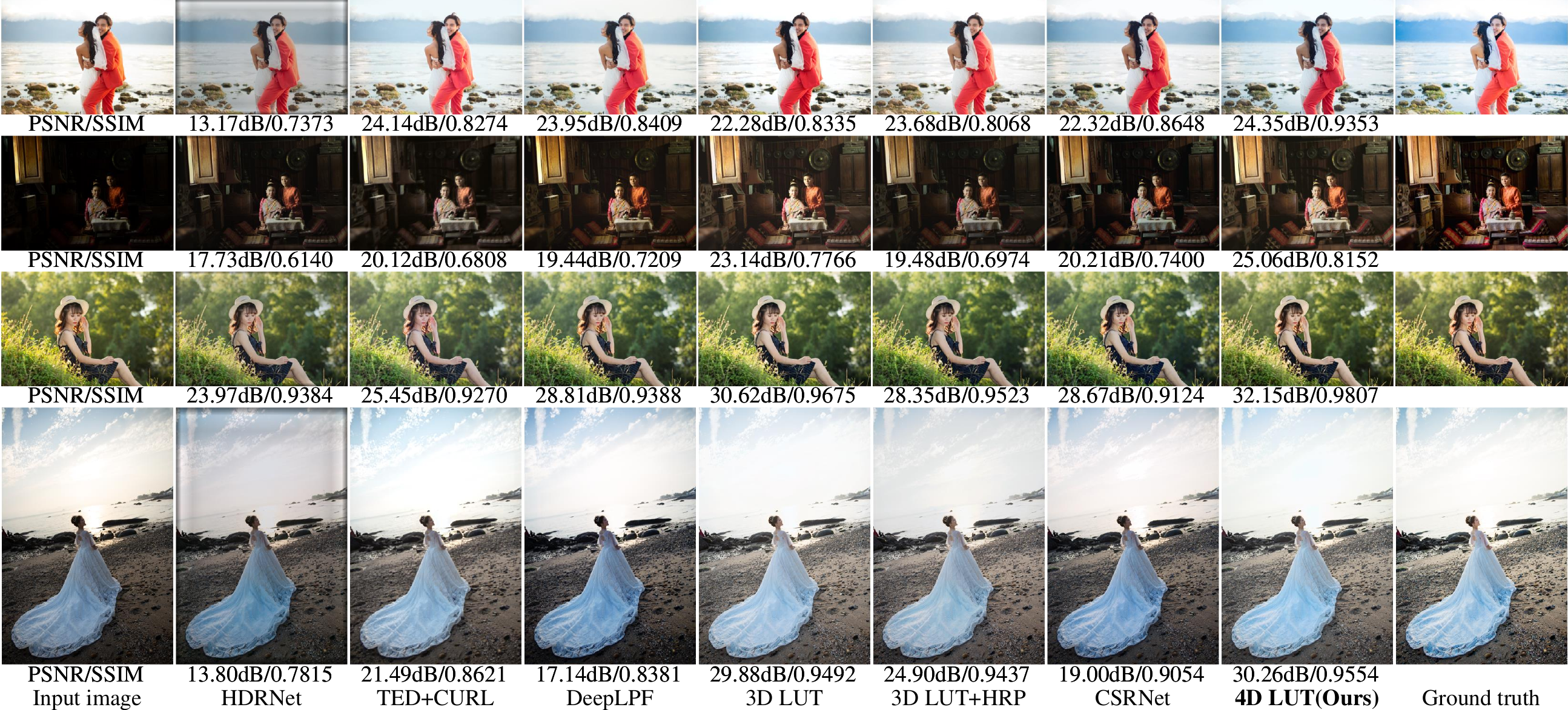}
  \caption{Visual comparison with state-of-the-arts on PPR10K-a~\cite{liang2021ppr10k} dataset. The quantitative comparison (PSNR$\uparrow$ and SSIM$\uparrow$) is shown at the bottom of each case.}
  \label{fig:case_ppr}
\end{figure*}

To further verify the generalization capabilities of 4D~LUT, we evaluate 4D~LUT on another larger-scale portrait photo retouching dataset PPR10K~\cite{liang2021ppr10k}. As shown in Tab.~\ref{table:3}, due to the well-designed 4D~LUT and the content-dependent learning capability, 4D~LUT achieves better results in all three experts-retouch results, which outperforms other SOTA methods between \textbf{0.22dB} to \textbf{0.30dB}. The performances demonstrate that 4D~LUT has strong generalization capabilities under different scenarios.

\subsubsection{Qualitative comparison}
To further compare the visual qualities of different algorithms, we show visual results enhanced by proposed 4D~LUT and other SOTA methods on different datasets in Fig.~\ref{fig:case_UPE} and Fig.~\ref{fig:case_ppr}. For fair comparisons, we either directly take the original enhanced results of the author-released or use author-released models to get results. 

It can be observed that 4D~LUT has a great improvement in visual quality and evaluation metrics (\emph{i.e.,}~PSNR and SSIM). For example, in the first row in Fig.~\ref{fig:case_ppr}, compared to other methods using a unified enhancer for landscapes and portraits, 4D~LUT can simultaneously obtain both blue landscapes and comfortably colored portraits. In the third row in Fig.~\ref{fig:case_ppr}, our 4D~LUT can enable a stronger enhancement capability by introducing an additional contextual dimension, which can produce brighter green plants and portraits. As the analysis mentioned above, the results verify that 4D~LUT has stronger enhancement capability and can achieve better results, especially for content-rich photos.

\subsubsection{Complexity analysis}
Model sizes and inference time are usually important in real applications. We follow previous works~\cite{he2020conditional,liu2022very} to report them enhancing an image with 360p by using an RTX 2080Ti GPU. As shown in Tab.~\ref{table:3}, compared with other SOTA methods, 4D~LUT achieves higher performance while keeping comparable \#Param. It should be emphasized that due to the expanded dimensionality, the parameters number of a 4D~LUT are larger compared to the 3DLUT~\cite{zeng2020learning,liang2021ppr10k} (\emph{i.e.,}~216K \emph{vs} 108K). Besides, 4D~LUT is slower compared to 3D~LUT~\cite{zeng2020learning} due to the generation of the context map and quadrilinear interpolation, but it also significantly exceeds the real-time runtime (\emph{i.e.,}~30fps). It should be emphasized that the quadrilinear interpolation of each pixel is independent of the others, and this transformation can be easily parallelized using the GPU, thus not adding much extra time.

\subsection{Ablation Study}
In this section, we conduct ablation experiments on model design and loss function on the MIT-Adobe-5K-UPE~\cite{wang2019underexposed} dataset.

\begin{table}[!t]
\begin{center}
\caption{Ablation study of each component in 4D~LUT. CE: context encoder. PE: parameter encoder.}
\label{table:4}
\begin{tabular}{ccc | c | c }
\hline

\hline
\multicolumn{3}{c|}{\textbf{Components}} &  \multirow{2}*{PSNR(dB)$\uparrow$} & \multirow{2}*{SSIM$\uparrow$} \\
Base   &       CE &     PE   &~ & ~  \\
\hline
\checkmark &       ~     & ~                   & 22.64&0.895   \\
\checkmark &  \checkmark & ~                   & 23.30& 0.897   \\
\checkmark &   ~  & \checkmark                & 24.65&0.920   \\
\checkmark & \checkmark  & \checkmark       & \textbf{24.96}& \textbf{0.924}  \\
\hline

\hline
\end{tabular}
\end{center}
\end{table}

\begin{figure}[!t]
\centering
\includegraphics[width=1.0\linewidth,page=4]{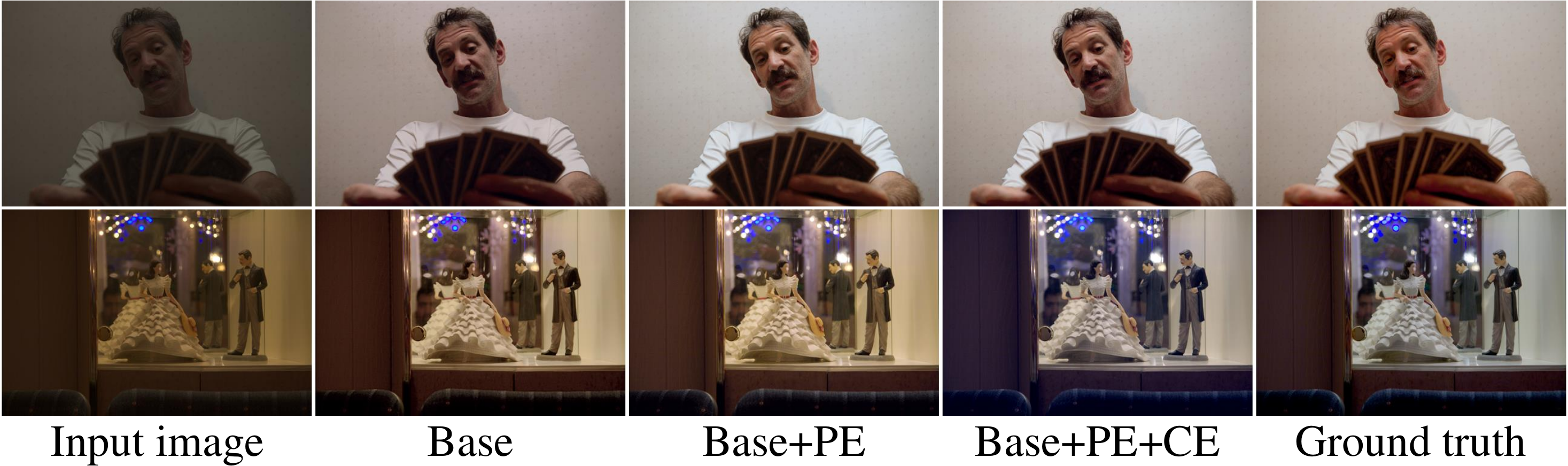}
\caption{Ablation study on the context encoder (CE) and parameter encoder (PE). (``4D~LUT'' can be interpreted as ``Base+PE+CE'').}
\label{fig:4}
\end{figure}

\begin{figure*}[!t]
\centering
\includegraphics[width=0.85\linewidth,page=5]{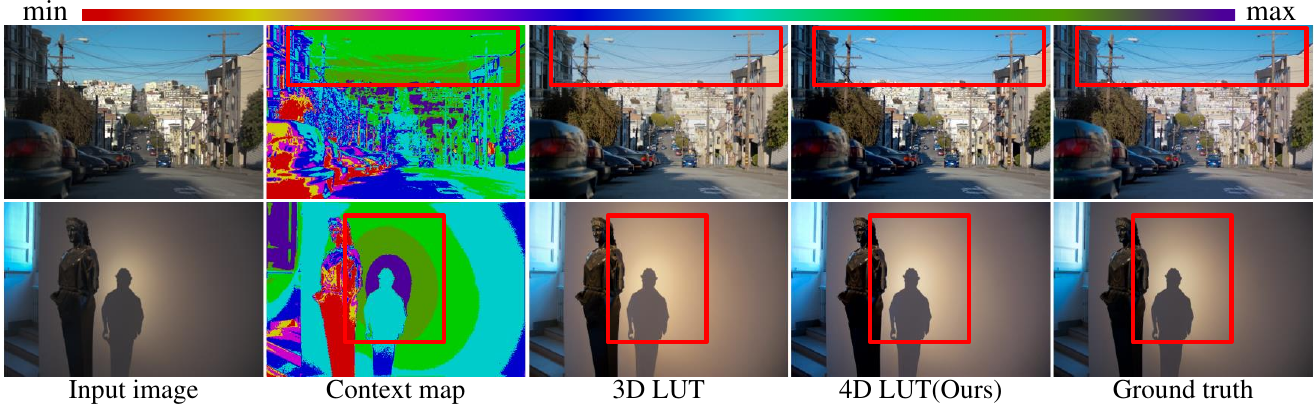}
\caption{Visualization of generated context map and enhanced results. The generated context map is visualized in eight kinds of colors from small to large according to the high-level semantic differences.}
\label{fig:5}
\end{figure*}

\subsubsection{Model design}
To demonstrate the effectiveness of each component in 4D~LUT, we conduct ablation experiments for each component. The experimental results are shown in Tab.~\ref{table:4}. The ``Base" indicates the result that no context encoder (\emph{i.e.,}~the context map is an all-zero image) and no parameter encoder (\emph{i.e.,}~directly summing the basis 4D~LUTs to fuse). ``CE" and ``PE" indicate the context encoder and parameter encoder, respectively.
The results show that the PSNR has increased by \textbf{1.34dB} by joining the CE. It demonstrates that the addition of the context encoder enables the network to learn content-dependent image enhancement, yielding stronger enhancement capabilities.
With the addition of PE, PSNR can be increased by \textbf{2.01dB}, which verifies that the learnable image-adaptive coefficients can be better fused into context-aware 4D~LUT, increasing the enhanced capability of the context-aware 4D~LUT.
When CE and PE are involved at the same time to boost each other, the performance is improved by \textbf{2.32dB}. 

We further explore the visual differences as shown in Fig.~\ref{fig:4}, context encoder can produce content-dependent image enhancement, while the parameter encoder produces richer color. It demonstrates the superiority of each component of 4D~LUT, which can get better performance for image enhancement.

\subsubsection{Loss function}
To demonstrate the effectiveness of each loss function in 4D~LUT, we conduct ablation experiments for them. 
The experimental results are shown in Tab.~\ref{table:5}. The ``$L_{r}$" indicates the pairwise reconstruction loss. ``$L_s$" and ``$L_m$" indicate the 4D smooth regularization loss and 4D monotonicity regularization loss, respectively.
With the addition of $L_s$, PSNR can be improved from 24.74dB to 24.79dB, which verifies that the 4D smooth regularization loss can ensure a stable transition from the input space (\emph{i.e.,}~RGBC) to the obtained color space (\emph{i.e.,}~RGB). 
When $L_m$ is involved, the color transformation can reserve the relative color brightness/saturation and cover the entire RGBC space, and the performance is improved to 24.96dB.
It demonstrates the superiority of each loss function of 4D~LUT, which can get better performance for image enhancement.

\begin{table}[!t]
\begin{center}
\caption{Ablation study of loss function used in 4D~LUT. CE: context encoder module. PE: parameter encoder module.}
\label{table:5}
\begin{tabular}{ccc | c | c }
\hline

\hline
\multicolumn{3}{c|}{\textbf{Loss function}} &  \multirow{2}*{PSNR(dB)$\uparrow$} & \multirow{2}*{SSIM$\uparrow$} \\
$L_r $  &       $L_s$ &     $L_m$   &~ & ~  \\
\hline
\checkmark &       ~     & ~                   & 24.74& 0.923   \\
\checkmark &  \checkmark & ~                   & 24.79&0.923   \\
\checkmark &   ~  & \checkmark                & 24.81& 0.924   \\
\checkmark & \checkmark  & \checkmark       & \textbf{24.96}& \textbf{0.924}  \\
\hline

\hline
\end{tabular}
\end{center}
\end{table}

\section{Discussions}
\label{Discussions}

In this section, to further demonstrate the reasonableness of 4D~LUT, we first visualize the proposed context map and context-aware 4D~LUT. Then we discuss the effect of bins $N_{bin}$ in the 4D~LUT and the number of basis 4D~LUT $N_{lut}$. Finally, we discuss the sensitivity of smooth regularization weight $\alpha_s$ and monotonicity regularization weight $\alpha_m$.

\begin{figure*}[!t]
\centering
\includegraphics[width=1.0\linewidth,page=6]{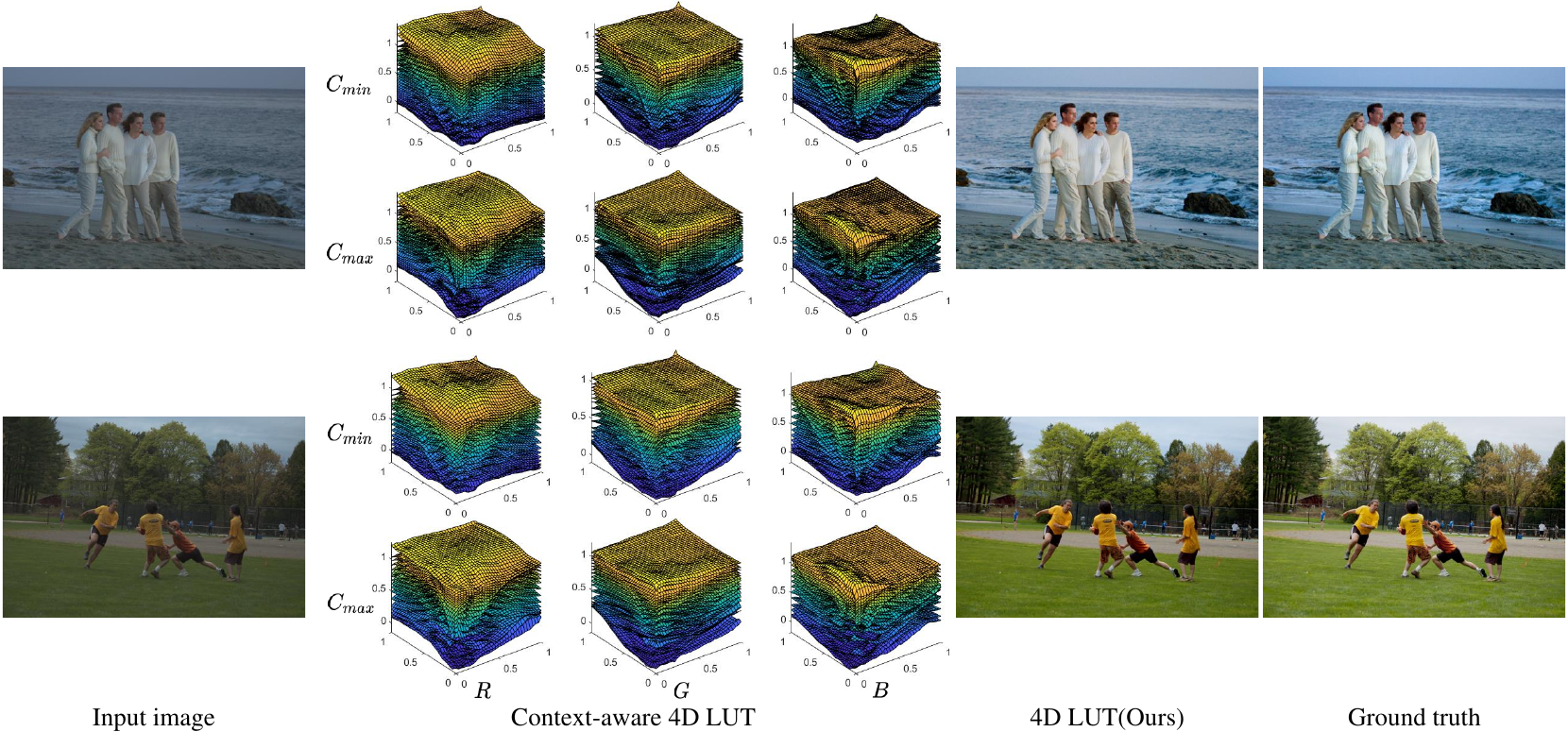}
\caption{Visualization of context-aware 4D~LUT and enhanced results. The context-aware 4D~LUT is visualized by dividing it into R, G, and B channels when the content values are minimum and maximum.}
\label{fig:6}
\end{figure*}

\subsection{Visualization of Context Map}
The context map is used to distinguish the high-level semantic differences between different regions. To explore the effectiveness of context map $C$ in 4D~LUT, we visualize it as shown in Fig.~\ref{fig:5}. Among them, we use eight kinds of colors to visualize the different contents in the generated context map from small to large, and it can be seen that the generated context map effectively divides the regions with different high-level semantic differences adaptively. Compared with the results of 3D~LUT not involving the context map, the results (indicated by the red box) demonstrate that our context-aware 4D~LUT is effective in achieving content-dependent enhancement with better results.

\subsection{Visualization of Context-aware 4D~LUT}
To better study this content-dependent enhancement property of context-aware 4D~LUT, in Fig.~\ref{fig:6}, we visualize the context-aware 4D~LUT for two different images and show their corresponding enhancement results. Besides, to visualize the differences of each of the $R$, $G$, and $B$ channels on the context-aware 4D~LUT more clearly, we fix the values of the context map to the maximum and minimum, respectively, and then visualize 17 slices (\emph{i.e.,}~\{1, 3, ..., 31, 33\}) of the whole 4D~LUT  (33 slices in total). 

As shown in Fig.~\ref{fig:6}, it is observed that for images containing blue ocean and green grass, our method can adaptively generate different 4D~LUT corresponding to different shapes. This demonstrates that 4D~LUT has the ability to establish color transformation relationships for images with different contents. Besides, the shapes of visualized LUT corresponding to different $C$-dimensions in each image also have significant differences. This proves that the 4D~LUT has the ability to perform different color transformations according to the different contents in each image, which enables finer control of color transformation and stronger image enhancement. It can be found that 4D~LUT can obtain pleasing visual results for different images or different contents in each image.

\subsection{Discussion on Number of Bins $N_{bin}$ in 4D~LUT}
To explore the influence of the number of bins $N_{bin}$ in 4D~LUT on the enhancement effect. As shown in Tab.~\ref{table:6}, we divide the 4D~LUT into different number of bins (\emph{i.e.,}~\{9, 17, 33, 64\}). 
As the value of $N_{bin}$ increases from 9 to 33, the PSNR is increased from 24.67dB to 24.96dB. It is because the increase in $N_{bin}$ makes the interpolated elements used for color transformation more accurate, and vice versa. Besides, to prevent the 4D~LUT from overfitting the color transformations of the training data and to preserve the generalization of the color transformations, we choose $N_{bin}$ as 33 in our experiments.

\begin{table}[!t]
\centering
\caption{Results of different number of bins $N_{bin}$ in 4D~LUT on MIT-Adobe-5K-UPE~\cite{wang2019underexposed} dataset.}
\label{table:6}
\begin{tabular}{c|c|c|c|c}
    \hline
    
    \hline
    \textbf{$N_{bin}$}     &   9  &  17  & 33  &  64\\
    \hline
    PSNR(dB)$\uparrow$  & 24.67 &24.79 & 24.96&25.03  \\
    \hline
    SSIM$\uparrow$   & 0.920 &0.923 &0.924&0.925  \\
    \hline
    
    \hline
\end{tabular}
\end{table}

\subsection{Discussion on Number of Basis 4D~LUTs $N_{lut}$}
To explore the influence of the number of basis 4D~LUTs $N_{lut}$ used on the enhancement effect.
As shown in Tab.~\ref{table:7}, we use different number of basis 4D~LUTs (\emph{i.e.,}~\{1, 2, 3, 4, 5\}) to fuse into a context-aware 4D~LUT that described in Sec.~\ref{4DLUTs Fusionzz}. The performance is positively correlated with the number of basis 4D~LUTs. It demonstrates using multiple basis 4D~LUTs improves the expressiveness of color transformations. Besides, it can be seen easily that when the number of base 4D~LUTs is increased from 1 to 3, its PSNR is significantly improved from 22.92dB to 24.96dB, while the improvement becomes smaller when the basis 4DLUTs is further increased. Therefore, by the trade-off between the number of model parameters and performance, we experimentally set $N_{lut}$ to 3.

\begin{table}[!t]
\centering
\caption{Results of different number of basis 4D~LUT $N_{lut}$ on MIT-Adobe-5K-UPE~\cite{wang2019underexposed} dataset.}
\label{table:7}
\begin{tabular}{c|c|c|c|c|c}
    \hline
    
    \hline
    \textbf{$N_{lut}$}     &   1  &  2  &  3  & 4  &  5\\
    \hline
    PSNR(dB)$\uparrow$  & 22.92 &24.27 & 24.96& 25.11 &25.16 \\
    \hline
    SSIM$\uparrow$   & 0.894 &0.915 & 0.924&0.925& 0.925  \\
    \hline
    
    \hline
\end{tabular}
\end{table}

\subsection{Discussion on Smooth Regularization Weight $\alpha_s$}
\begin{figure}[!t]
\centering
\includegraphics[width=1.0\linewidth,page=7]{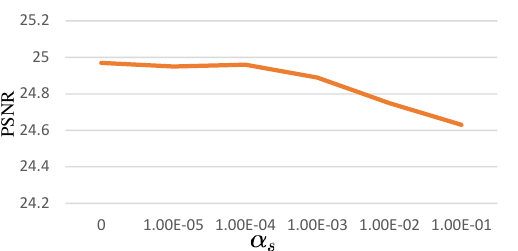}
\caption{Sensitivity of smooth regularization weight $\alpha_s$.}
\label{fig:7}
\end{figure}

As described in Sec.~\ref{4D smooth regularization} above, we use a 4D smooth regularization to ensure the stability of the locally color transformation. Therefore, we perform experiments by setting the smooth regularization weight $\alpha_s$ distributed in $\{0, e^{-5}, e^{-4}, e^{-3}, e^{-2}, e^{-1}\}$ to select the appropriate values in Equ.~\ref{lossfunctionzz}. As shown in Fig.~\ref{fig:7}, it can be seen that when $\alpha_s$ is too large, excessive smoothing makes A missing a detailed description of the color transformations, while reducing performance. On the contrary, when $\alpha_s$ is too small, insufficient smoothing makes the network lack the ability to generalize the color transformations. We set $\alpha_s$ as $0.0001$ in the final experimentally.

\subsection{Discussion on Monotonicity Regularization Weight $\alpha_m$}
\begin{figure}[!t]
\centering
\includegraphics[width=1.0\linewidth,page=8]{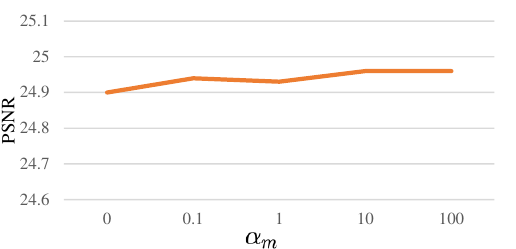}
\caption{Sensitivity of monotonicity regularization weight $\alpha_m$.}
\label{fig:8}
\end{figure}

To explore the influence of the monotonicity regularization weight on the color enhancement effect in Sec.~\ref{4D monotonicity regularization}. We perform experiments by setting the smooth regularization weight $\alpha_m$ distributed in $\{0, 0.1, 1, 10, 100\}$ in Equ.~\ref{lossfunctionzz}. The experimental results are shown in Fig.~\ref{fig:8}. Compared with not adding the monotonicity regularization, the monotonicity regularization of the 4D~LUT can preserve the relative color brightness, and make the color transformation cover the whole RGBC space. The impact of the larger monotonicity regularization weight $\alpha_m$ is minor, and we experimentally set $\alpha_m$ to 10 finally.

\section{Conclusion}
\label{Conclusion}
In this paper, we extend the lookup table architecture into a 4-dimensional space and propose a novel learnable context-aware 4-dimensional lookup table (4D~LUT).
It includes four closely-related components. 1) A context encoder is used to generate the content-dependent context map. 2) A parameter encoder for generating image-adaptive coefficients. 3) A 4D~LUTs fusion module integrates the coefficients and learnable basis 4D~LUTs into a content-aware 4D~LUT. 4) A quadrilinear interpolation module output the enhanced image.
This design introduces the image content and enables finer control of color transformations for pixels in each image, resulting in content-dependent image enhancement via learning image contents adaptively.
Experimental results show significantly superior between the proposed 4D~LUT and existing SOTA models.
In the future, we will focus on extending our 4D~LUT in more low-level vision tasks through more explorations.

\bibliographystyle{plain}
\bibliography{ref}

\begin{thebibliography}{10}

\bibitem{aubry2014fast}
Mathieu Aubry, Sylvain Paris, Samuel~W Hasinoff, Jan Kautz, and Fr{\'e}do
  Durand.
\newblock Fast local laplacian filters: Theory and applications.
\newblock {\em ACM TOG}, 33(5):1--14, 2014.

\bibitem{bychkovsky2011learning}
Vladimir Bychkovsky, Sylvain Paris, Eric Chan, and Fr{\'e}do Durand.
\newblock Learning photographic global tonal adjustment with a database of
  input/output image pairs.
\newblock In {\em CVPR 2011}, pages 97--104. IEEE, 2011.

\bibitem{chen2017photographic}
Qifeng Chen and Vladlen Koltun.
\newblock Photographic image synthesis with cascaded refinement networks.
\newblock In {\em ICCV}, pages 1511--1520, 2017.

\bibitem{chen2018deep}
Yu-Sheng Chen, Yu-Ching Wang, Man-Hsin Kao, and Yung-Yu Chuang.
\newblock Deep photo enhancer: Unpaired learning for image enhancement from
  photographs with {GANs}.
\newblock In {\em CVPR}, pages 6306--6314, 2018.

\bibitem{gharbi2017deep}
Micha{\"e}l Gharbi, Jiawen Chen, Jonathan~T Barron, Samuel~W Hasinoff, and
  Fr{\'e}do Durand.
\newblock Deep bilateral learning for real-time image enhancement.
\newblock {\em ACM TOG}, 36(4):1--12, 2017.

\bibitem{goodfellow2014generative}
Ian Goodfellow, Jean Pouget-Abadie, Mehdi Mirza, Bing Xu, David Warde-Farley,
  Sherjil Ozair, Aaron Courville, and Yoshua Bengio.
\newblock Generative adversarial nets.
\newblock {\em NeurIPS}, 27, 2014.

\bibitem{guo2020zero}
Chunle Guo, Chongyi Li, Jichang Guo, Chen~Change Loy, Junhui Hou, Sam Kwong,
  and Runmin Cong.
\newblock Zero-reference deep curve estimation for low-light image enhancement.
\newblock In {\em CVPR}, pages 1780--1789, 2020.

\bibitem{guo2016lime}
Xiaojie Guo, Yu~Li, and Haibin Ling.
\newblock {LIME}: Low-light image enhancement via illumination map estimation.
\newblock {\em IEEE TIP}, 26(2):982--993, 2016.

\bibitem{he2020conditional}
Jingwen He, Yihao Liu, Yu~Qiao, and Chao Dong.
\newblock Conditional sequential modulation for efficient global image
  retouching.
\newblock In {\em ECCV}, pages 679--695. Springer, 2020.

\bibitem{he2012guided}
Kaiming He, Jian Sun, and Xiaoou Tang.
\newblock Guided image filtering.
\newblock {\em IEEE TPAMI}, 35(6):1397--1409, 2012.

\bibitem{he2016deep}
Kaiming He, Xiangyu Zhang, Shaoqing Ren, and Jian Sun.
\newblock Deep residual learning for image recognition.
\newblock In {\em CVPR}, pages 770--778, 2016.

\bibitem{hu2018exposure}
Yuanming Hu, Hao He, Chenxi Xu, Baoyuan Wang, and Stephen Lin.
\newblock Exposure: A white-box photo post-processing framework.
\newblock {\em ACM TOG}, 37(2):1--17, 2018.

\bibitem{ignatov2017dslr}
Andrey Ignatov, Nikolay Kobyshev, Radu Timofte, Kenneth Vanhoey, and Luc
  Van~Gool.
\newblock {DSLR}-quality photos on mobile devices with deep convolutional
  networks.
\newblock In {\em ICCV}, pages 3277--3285, 2017.

\bibitem{isola2017image}
Phillip Isola, Jun-Yan Zhu, Tinghui Zhou, and Alexei~A Efros.
\newblock Image-to-image translation with conditional adversarial networks.
\newblock In {\em CVPR}, pages 1125--1134, 2017.

\bibitem{kim2020global}
Han-Ul Kim, Young~Jun Koh, and Chang-Su Kim.
\newblock Global and local enhancement networks for paired and unpaired image
  enhancement.
\newblock In {\em ECCV}, pages 339--354. Springer, 2020.

\bibitem{kim2020pienet}
Han-Ul Kim, Young~Jun Koh, and Chang-Su Kim.
\newblock {PieNet}: Personalized image enhancement network.
\newblock In {\em ECCV}, pages 374--390. Springer, 2020.

\bibitem{kingma2014adam}
Diederik~P Kingma and Jimmy Ba.
\newblock Adam: A method for stochastic optimization.
\newblock In {\em ICLR}, 2015.

\bibitem{kneubuehler2020flexible}
Dario Kneubuehler, Shuhang Gu, Luc~Van Gool, and Radu Timofte.
\newblock Flexible example-based image enhancement with task adaptive global
  feature self-guided network.
\newblock In {\em ECCV}, pages 343--358. Springer, 2020.

\bibitem{kosugi2020unpaired}
Satoshi Kosugi and Toshihiko Yamasaki.
\newblock Unpaired image enhancement featuring
  reinforcement-learning-controlled image editing software.
\newblock In {\em AAAI}, volume~34, pages 11296--11303, 2020.

\bibitem{land1977retinex}
Edwin~H Land.
\newblock The retinex theory of color vision.
\newblock {\em Scientific american}, 237(6):108--129, 1977.

\bibitem{li2021low}
Chongyi Li, Chunle Guo, Ling-Hao Han, Jun Jiang, Ming-Ming Cheng, Jinwei Gu,
  and Chen~Change Loy.
\newblock Low-light image and video enhancement using deep learning: A survey.
\newblock {\em IEEE TPAMI}, (01):1--1, 2021.

\bibitem{li2018structure}
Mading Li, Jiaying Liu, Wenhan Yang, Xiaoyan Sun, and Zongming Guo.
\newblock Structure-revealing low-light image enhancement via robust retinex
  model.
\newblock {\em IEEE TIP}, 27(6):2828--2841, 2018.

\bibitem{liang2021ppr10k}
Jie Liang, Hui Zeng, Miaomiao Cui, Xuansong Xie, and Lei Zhang.
\newblock {PPR10K}: A large-scale portrait photo retouching dataset with
  human-region mask and group-level consistency.
\newblock In {\em CVPR}, pages 653--661, 2021.

\bibitem{liang2021cameranet}
Zhetong Liang, Jianrui Cai, Zisheng Cao, and Lei Zhang.
\newblock {CameraNet}: A two-stage framework for effective camera isp learning.
\newblock {\em IEEE TIP}, 30:2248--2262, 2021.

\bibitem{liu2017unsupervised}
Ming-Yu Liu, Thomas Breuel, and Jan Kautz.
\newblock Unsupervised image-to-image translation networks.
\newblock {\em NeurIPS}, 30, 2017.

\bibitem{liu2022very}
Yihao Liu, Jingwen He, Xiangyu Chen, Zhengwen Zhang, Hengyuan Zhao, Chao Dong,
  and Yu~Qiao.
\newblock Very lightweight photo retouching network with conditional sequential
  modulation.
\newblock {\em IEEE TMM}, 2022.

\bibitem{lu2015spatio}
Shao-Ping Lu, Beerend Ceulemans, Adrian Munteanu, and Peter Schelkens.
\newblock Spatio-temporally consistent color and structure optimization for
  multiview video color correction.
\newblock {\em IEEE TMM}, 17(5):577--590, 2015.

\bibitem{ma2022toward}
Long Ma, Tengyu Ma, Risheng Liu, Xin Fan, and Zhongxuan Luo.
\newblock Toward fast, flexible, and robust low-light image enhancement.
\newblock In {\em CVPR}, pages 5637--5646, 2022.

\bibitem{moran2020deeplpf}
Sean Moran, Pierre Marza, Steven McDonagh, Sarah Parisot, and Gregory Slabaugh.
\newblock {DeepLPF}: Deep local parametric filters for image enhancement.
\newblock In {\em Proceedings of the IEEE/CVF conference on computer vision and
  pattern recognition}, pages 12826--12835, 2020.

\bibitem{moran2021curl}
Sean Moran, Steven McDonagh, and Gregory Slabaugh.
\newblock {CURL}: Neural curve layers for global image enhancement.
\newblock In {\em 2020 25th International Conference on Pattern Recognition
  (ICPR)}, pages 9796--9803. IEEE, 2021.

\bibitem{pan2021miegan}
Zhaoqing Pan, Feng Yuan, Jianjun Lei, Wanqing Li, Nam Ling, and Sam Kwong.
\newblock {MIEGAN}: mobile image enhancement via a multi-module cascade neural
  network.
\newblock {\em IEEE TMM}, 24:519--533, 2021.

\bibitem{park2018distort}
Jongchan Park, Joon-Young Lee, Donggeun Yoo, and In~So Kweon.
\newblock {Distort-and-Recover}: Color enhancement using deep reinforcement
  learning.
\newblock In {\em CVPR}, pages 5928--5936, 2018.

\bibitem{qi2021comprehensive}
Yunliang Qi, Zhen Yang, Wenhao Sun, Meng Lou, Jing Lian, Wenwei Zhao, Xiangyu
  Deng, and Yide Ma.
\newblock A comprehensive overview of image enhancement techniques.
\newblock {\em Archives of Computational Methods in Engineering}, pages 1--25,
  2021.

\bibitem{ronneberger2015u}
Olaf Ronneberger, Philipp Fischer, and Thomas Brox.
\newblock {U-Net}: Convolutional networks for biomedical image segmentation.
\newblock In {\em International Conference on Medical image computing and
  computer-assisted intervention}, pages 234--241. Springer, 2015.

\bibitem{schwartz2018deepisp}
Eli Schwartz, Raja Giryes, and Alex~M Bronstein.
\newblock {DeepISP}: Toward learning an end-to-end image processing pipeline.
\newblock {\em IEEE TIP}, 28(2):912--923, 2018.

\bibitem{tomasi1998bilateral}
Carlo Tomasi and Roberto Manduchi.
\newblock Bilateral filtering for gray and color images.
\newblock In {\em Sixth international conference on computer vision (IEEE Cat.
  No. 98CH36271)}, pages 839--846. IEEE, 1998.

\bibitem{wang2019underexposed}
Ruixing Wang, Qing Zhang, Chi-Wing Fu, Xiaoyong Shen, Wei-Shi Zheng, and Jiaya
  Jia.
\newblock Underexposed photo enhancement using deep illumination estimation.
\newblock In {\em CVPR}, pages 6849--6857, 2019.

\bibitem{wang2021real}
Tao Wang, Yong Li, Jingyang Peng, Yipeng Ma, Xian Wang, Fenglong Song, and
  Youliang Yan.
\newblock Real-time image enhancer via learnable spatial-aware 3d lookup
  tables.
\newblock In {\em CVPR}, pages 2471--2480, 2021.

\bibitem{wang2004image}
Zhou Wang, Alan~C Bovik, Hamid~R Sheikh, and Eero~P Simoncelli.
\newblock Image quality assessment: from error visibility to structural
  similarity.
\newblock {\em IEEE TIP}, 13(4):600--612, 2004.

\bibitem{wei2018deep}
Chen Wei, Wenjing Wang, Wenhan Yang, and Jiaying Liu.
\newblock Deep retinex decomposition for low-light enhancement.
\newblock In {\em BMVC}, 2018.

\bibitem{xu2013generalized}
Hongteng Xu, Guangtao Zhai, Xiaolin Wu, and Xiaokang Yang.
\newblock Generalized equalization model for image enhancement.
\newblock {\em IEEE TMM}, 16(1):68--82, 2013.

\bibitem{yan2016automatic}
Zhicheng Yan, Hao Zhang, Baoyuan Wang, Sylvain Paris, and Yizhou Yu.
\newblock Automatic photo adjustment using deep neural networks.
\newblock {\em ACM TOG}, 35(2):1--15, 2016.

\bibitem{yang2022adaint}
Canqian Yang, Meiguang Jin, Xu~Jia, Yi~Xu, and Ying Chen.
\newblock {AdaInt}: Learning adaptive intervals for {3D} lookup tables on
  real-time image enhancement.
\newblock In {\em CVPR}, pages 17522--17531, 2022.

\bibitem{yu2018deepexposure}
Runsheng Yu, Wenyu Liu, Yasen Zhang, Zhi Qu, Deli Zhao, and Bo~Zhang.
\newblock {DeepExposure}: Learning to expose photos with asynchronously
  reinforced adversarial learning.
\newblock {\em NeurIPS}, 31, 2018.

\bibitem{zamir2020learning}
Syed~Waqas Zamir, Aditya Arora, Salman Khan, Munawar Hayat, Fahad~Shahbaz Khan,
  Ming-Hsuan Yang, and Ling Shao.
\newblock Learning enriched features for real image restoration and
  enhancement.
\newblock In {\em ECCV}, pages 492--511. Springer, 2020.

\bibitem{zeng2020learning}
Hui Zeng, Jianrui Cai, Lida Li, Zisheng Cao, and Lei Zhang.
\newblock Learning image-adaptive {3D} lookup tables for high performance photo
  enhancement in real-time.
\newblock {\em IEEE TPAMI}, 2020.

\bibitem{zhu2017unpaired}
Jun-Yan Zhu, Taesung Park, Phillip Isola, and Alexei~A Efros.
\newblock Unpaired image-to-image translation using cycle-consistent
  adversarial networks.
\newblock In {\em ICCV}, pages 2223--2232, 2017.

\end{thebibliography}
\end{document}